\documentclass[10pt,journal, onecolumn]{IEEEtran} 

\IEEEoverridecommandlockouts

\usepackage{cite}
\usepackage{amsmath,amssymb,amsfonts}
\usepackage{graphicx}
\usepackage{textcomp}
\usepackage{xcolor}
\usepackage{siunitx}
\usepackage{booktabs}
\usepackage{float}
\usepackage{dblfloatfix}
\usepackage{epsfig}
\usepackage{caption}
\usepackage{float}
\usepackage{url}

\usepackage{algorithm}
\usepackage[noend]{algpseudocode}

\usepackage{todonotes}

\usepackage{soul}

\usepackage[utf8]{inputenc}
\usepackage{booktabs,caption}
\usepackage[flushleft]{threeparttable}
\usepackage{multirow}
\usepackage{dblfloatfix}
\usepackage{comment}
\newcommand{\red}{\color{black}}
\newcommand{\rd}{\color{black}}

\def\BibTeX{{\rm B\kern-.05em{\sc i\kern-.025em b}\kern-.08em
    T\kern-.1667em\lower.7ex\hbox{E}\kern-.125emX}}

\begin{document}

\title{A $334\mu W$ $0.158mm^2$ Saber Learning with Rounding based Post-Quantum Crypto Accelerator}
\title{A $334\mu W$ $0.158mm^2$ ASIC for Post-Quantum Key-Encapsulation Mechanism Saber with Low-latency Striding Toom-Cook multiplication\\ \large{Authors' version}}
\title{A $334\mu W$ $0.158mm^2$ ASIC for Post-Quantum Key-Encapsulation Mechanism Saber with Low-latency Striding Toom-Cook Multiplication Authors Version}




\author{\IEEEauthorblockA{Archisman Ghosh\IEEEauthorrefmark{1},
Jose Maria Bermudo Mera\IEEEauthorrefmark{3}\IEEEauthorrefmark{4},
Angshuman Karmakar\IEEEauthorrefmark{3}\IEEEauthorrefmark{5},
Debayan Das\IEEEauthorrefmark{1},
Santosh Ghosh\IEEEauthorrefmark{2},
Ingrid Verbauwhede\IEEEauthorrefmark{3},
and~Shreyas~Sen\IEEEauthorrefmark{1}} \\
\IEEEauthorblockA{\IEEEauthorrefmark{1}School of Electrical and Computer Engineering,
Purdue University, West Lafayette, IN, USA} 
\IEEEauthorblockA{\IEEEauthorrefmark{3}COSIC, KU Leuven, Belgium}
\IEEEauthorblockA{\IEEEauthorrefmark{2}Intel Labs, Intel Corporation, Hillsboro, OR, USA}
\IEEEauthorblockA{\IEEEauthorrefmark{4}PQShield Ltd, UK}
\IEEEauthorblockA{\IEEEauthorrefmark{5}Indian Institute of Technology Kanpur, India}
}

\maketitle
\vspace{-2mm}
\begin{abstract}
The hard mathematical problems that assure the security of our current public-key cryptography (RSA, ECC) are broken if and when a quantum computer appears rendering them ineffective for use in the quantum era. 
Lattice based cryptography is a novel approach to public key cryptography, of which the mathematical investigation (so far) resists attacks from quantum computers. By choosing a module learning with errors (MLWE) algorithm as the next standard, National Institute of Standard \& Technology (NIST) follows this approach. The multiplication of polynomials is the central bottleneck in the computation of lattice based cryptography. Because public key cryptography is mostly used to establish common secret keys, focus is on compact area, power and energy budget and to a lesser extent on throughput or latency. While most other work focuses on optimizing number theoretic transform (NTT) based multiplications, in this paper we highly optimize a Toom-Cook based multiplier. We demonstrate that a memory-efficient striding Toom-Cook with lazy interpolation, results in a highly compact, low power implementation, which on top enables a very regular memory access scheme. To demonstrate the efficiency, we integrate this multiplier into a Saber post-quantum accelerator, one of the four NIST finalists. Algorithmic innovation to reduce active memory, timely clock gating and shift-add multiplier has helped to achieve 38\% less power than state-of-the art PQC core, 4 $\times$ less memory, 36.8\% reduction in multiplier energy and 118$\times$ reduction in active power with respect to state-of-the-art Saber accelerator (not silicon verified). This accelerator consumes $0.158mm^2$ active area which is lowest reported till date despite process disadvantages of the state-of-the-art designs. 


%



\end{abstract}


\IEEEpeerreviewmaketitle

\section{Introduction}\label{sec:intro}
In the past few decades, we have witnessed an unparalleled expansion of digital technologies in to all parts of our life. Considering the huge convenience and cost-effectiveness facilitated by these technologies in different sectors such as commerce, education, banking, communications, legality, etc, this penetration is expected to grow further in the future. However, if we look closely such proliferation of these technologies would never have been possible if there were no \emph{assurances} such as the ability to verify the identity of the other communicating entity, mechanisms to prevent and detect tampering of data, the assertion of privacy between two or more communicating entities, methods to store one's secret and sensitive data, etc, while using these technologies. Cryptography is the discipline that studies different techniques to provide such assurance. Cryptography primarily emanated from the need for the secret exchange of information between allies and friendly entities in a hostile environment and until the end of the second world war, cryptography found little use outside military applications. Nevertheless, with the advancement in the field of computer science and the increase in digital technologies the field of cryptography flourished and ramified to provide solutions for many different problems. 

Broadly, cryptography can be classified into two ways, symmetric-key cryptography and asymmetric-key or public-key cryptography (PKC). Symmetric-key cryptographic schemes are fast, low resource-hungry and can be used to encrypt data in bulk. However, they assume that a common \emph{secret-key} has been shared among communicating parties beforehand. This requirement is often difficult to satisfy on real-world applications. On the other hand, PKC schemes, do not have such pre-requisite of shared secret-keys among communicating parties.
Although they are typically slower, more resource-hungry, and can encrypt smaller data relative to the symmetric-key cryptographic schemes. Due to this complementary nature, symmetric-key and asymmetric-key cryptographic schemes are often used in tandem in many real-world applications. For example, in transport layer security (TLS)~\cite{TLS} protocol PKC schemes or specifically key-exchange schemes are used to establish the secret session key which is then used to encrypt ensuing communication using symmetric-key schemes. Another example is blockchain which is primarily constructed by combining digital-signatures~\cite{digital_signature} (asymmetric-key) and hash functions~\cite{hash_functions} (symmetric-key). The security assurances of both of these schemes arise from the computational intractability of some underlying hard problems i.e. it is infeasible to solve even one instance of such problems with practical parameters using the best-known algorithms to solve them which are running on a classical computer. However, in the context of quantum computers, this assumption does not always hold. For example, large integer-factorization and elliptic-curve discrete logarithm problems are two such hard problems that are used to construct two very well-known PKC schemes Rivest-Shamir-Adleman (RSA)~\cite{rsa} and elliptic-curve cryptography (ECC)~\cite{koblitz, miller} respectively. However, there are quantum algorithms such as Shor's~\cite{Shor} and Proos-Zalka's~\cite{proos_zalka} algorithm that can solve these two problems \emph{easily} in polynomial time using a \emph{large} quantum computer. Considering that our current public-key infrastructure is based mostly on RSA- and ECC-based schemes, we need to replace these schemes with quantum-resistant PKC to maintain the security of our digital world in the future. For symmetric-key, this problem is much less severe. The best-known quantum algorithm that can be used against symmetric-key cryptography is Grover's search~\cite{grover_search} algorithm. This algorithm gives a quadratic speed-up in searching an unordered list. Hence, the adverse effect of Grover's search algorithm can be easily overcome by doubling the key lengths of symmetric-key cryptography.

\begin{figure}[!ht]
  \centering
  \includegraphics[scale=.8]{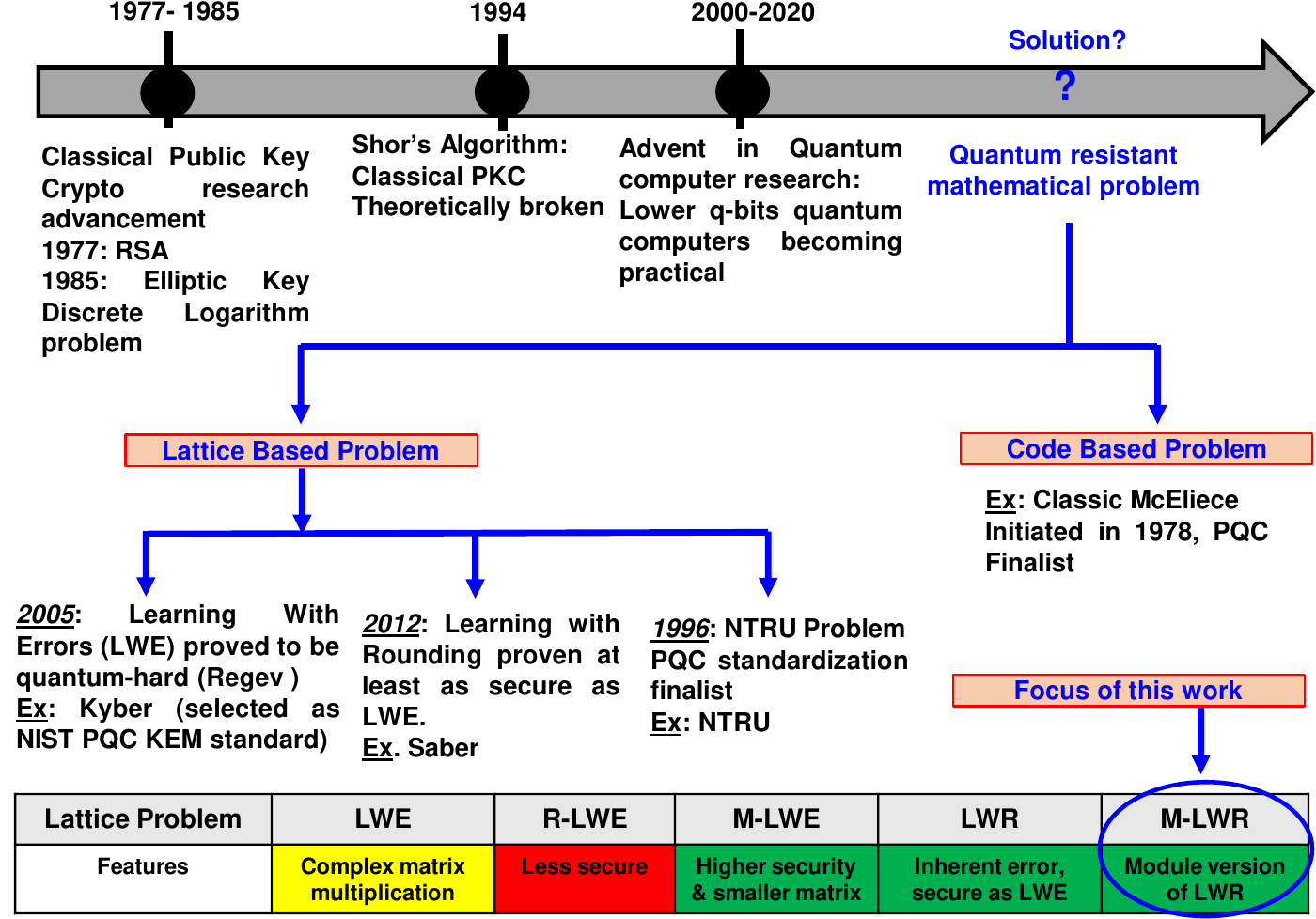}
  \caption{Timeline of PKC. Quantum computing research has motivated the use of quantum-hard mathematical problems.}
  \label{timeline}
\end{figure}

Post-quantum cryptography (PQC) studies hard problems that remain hard to solve even in the presence of large quantum computers. The post-quantum standardization initiative by the National Institute of Standards and Technology (NIST)~\cite{nist_PQC} in 2017 was a prudent step towards developing PQC. Initially, $59$ and $23$ schemes were submitted in key-encapsulation mechanism (KEM) and digital signature categories, respectively. 3 lattice-based (Saber, Kyber, and NTRU) and 1 code-based algorithm (Classic McEliece) were selected after three rounds of research and analysis by NIST for final consideration. 

{\rd A historic timeline of PKC research along with the pros and cons of different lattice constructions is summarized in Figure~\ref{timeline}. NIST has recently mentioned Kyber as the new standard. However, the NIST report on finalist candidates~\cite{nist_third_round_report} mentioned that all the different experiments suggest Saber has strong security at par with Kyber. }

\subsection{Lattice-based Cryptography}
The standard hard problem used to build {\rd lattice-based cryptographic {(LBC)}} schemes is the Learning with errors (LWE) problem~\cite{DBLP:journals/jacm/Regev04}. It states that it is hard to distinguish between $n$ uniformly random samples $a\in \mathbb{Z}_q^n$ and $n$ samples drawn as $(a_i,b_i=\left \langle a_i\cdot s \right \rangle+e_i)$ without having information about the secret $s$ or the random error terms $e_i$. The LWE problem can be used to build public-key cryptography considering the matrix $A$ and vector $b$ (formed by the samples $a_i$ and $b_i$, respectively) as the public key and $s$ as the secret key. The main drawback of LWE schemes is the expensive matrix-vector multiplication $A\cdot s$ with a large range $n$.

The variant ring-Learning with errors (R-LWE)~\cite{DBLP:conf/eurocrypt/LyubashevskyPR10} was proposed to improve the efficiency of LWE schemes. In R-LWE, there is a single sample $(a,t=a\ast s+e)\in R_q\times R_q$ where each of the elements is now a polynomial belonging to the ring $R_q=\mathbb{Z}_q[x]/(x^n+1)$. R-LWE can be connected to LWE by looking at the new public matrix $A$ as a matrix formed by rotations of the polynomial $a$. However, the central operation is now a polynomial convolution for which there exist algorithms with better time complexity.

While R-LWE schemes are more efficient, there is less trust in their security due to the extra algebraic structure conferred to the mathematical problem, which might be exploitable by an attacker. Module Learning with errors (M-LWE)~\cite{DBLP:journals/dcc/LangloisS15} was proposed as an intermediate solution between LWE and R-LWE. In M-LWE the public matrix $A$ has a much smaller dimension than in LWE, but each of the elements is a polynomial rather than an integer. The secret $s$ and the error term $e$ are also short vectors of polynomials. It offers higher security confidence than R-LWE while still benefiting from the better efficiency of polynomial arithmetic.


{\rd
In this work, we work with Saber key-encapsulation mechanism{~\cite{saber}} which is based on the hard problem known as Learning with rounding (LWR). LWR is a variant of the aforementioned LWE where the error term is introduced implicitly by \emph{rounding} unlike the explicit addition of error terms in LWE. Therefore, for a given matrix $A\in\mathbb{Z}_q^{m\times n}$, random secret and errors $s,e\in\mathbb{Z}_q^n$, the LWE and LWR samples are given as $(A, b=A\cdot s+e)$ and $(A, b=\lfloor A\cdot s\rceil)$ respectively. Here, we want to point that $A\cdot s$ is a very good \emph{entropy-diffuser}. In fact, $A\cdot s$ itself can be used as a pseudo-random function. However, since $A$ is invertible with a very high probability, given the samples $b=A\cdot s$, the secret can be recovered trivially. Therefore, this cannot be used in cryptography. In Regev's{~\cite{DBLP:journals/jacm/Regev04}} seminal LWE paper, it was shown that even after adding small error terms $e$ the samples $b=A\cdot s +e$ remain computationally indistinguishable from uniformly generated random samples $\mathbf{u}$ \textit{i.e.} $\mathbf{u} \overset{\mathrm{comp}}{\approx} b$. This removes the possibility of any correlation attack on LWE. In fact, the decisional problem of LWE \textit{i.e.} distinguishing samples $b$ from $u$ can be shown to be equivalent to the computational LWE problem mentioned above{~\cite{DBLP:journals/jacm/Regev04}}. In his paper, Regev also showed that given samples $b$, recovering the secret $s$ is at least as hard as solving GAP shortest vector problems (SVP) in random lattices in the worst case \cite{peikert2009public}. As no known quantum algorithms can solve GapSVP problem with an overwhelming advantage over their classical counterparts, this reduction from GapSVP to LWE engenders the quantum hardness of LWE-based schemes.

Banerjee et al.{~\cite{DBLP:conf/eurocrypt/BanerjeePR12}} first introduced the LWR problem which removes the necessity of adding the explicit error terms and introducing the error implicitly by \emph{chopping} the lower order bits or rounding to a smaller field $\mathbb{Z}_p$. They showed that the LWR problem is at least as hard as LWE problem. One of their main results was to show that if $q/p$ is sufficiently big then the $\lfloor A\cdot s\rceil\overset{stat}{\approx}\lfloor A\cdot s+e\rceil$. So combining this with the Regev's{~\cite{DBLP:journals/jacm/Regev04,DBLP:conf/eurocrypt/LyubashevskyPR10}} result mentioned earlier, one can show that 
$\lfloor A\cdot s\rceil\overset{\mathrm{stat}}{\approx}\lfloor A\cdot s+e\rceil\overset{\mathrm{comp}}{\approx}\lfloor \mathbf{u}\rceil$. Therefore, as before it is difficult to distinguish between LWR and uniform random distribution. This implies that the correlation between the LWR sample and the secret is not more than the correlation between an LWE sample and its respective secret. Later works by Bogdanov et al.{~\cite{DBLP:conf/tcc/BogdanovGMRR16}}, Alperin-Sheriff et al.{~\cite{DBLP:journals/iacr/Alperin-Sheriff16}}, and Alwen et al.{~\cite{DBLP:conf/crypto/AlwenKPW13}} further reduced the required value of $q/p$ for these reductions to be held. Quite unfortunately, delving deeper into the security analysis of LWE or LWR is out of the scope of this work. We kindly refer interested readers to the original articles for more details. For a detailed discussion on deriving the security of Saber from LWR or specifically Module-LWR, we refer to the NIST specification document{~\cite{saber}} of Saber submission.
}

\subsection{Saber}

Saber~\cite{saber} is a lattice-based post-quantum KEM. It is one of the $4$ finalist schemes of the NIST standardization procedure in the KEM category. Therefore, it has gone through extensive and rigorous scrutiny by the cryptographic community. This is a testimony of Saber for efficiency, theoretical security, and resilience to physical attacks. 


Saber's security relies on the hardness of solving the module Learning with rounding (\textbf{M-LWR}) problem. \textbf{Key generation}, described in Algorithm~\ref{alg:KeyGenPKE}, starts by generating a truly random $256$-bit seed. This seed is expanded with a function based on the extendable output function SHAKE-128 to generate the public matrix $A$. A similar approach is followed to generate the secret $s$, but the coefficients of $s$ follow a discrete binomial distribution $\beta_{\mu}$ with parameter $\mu$ rather than being uniformly distributed. The sample $b$ is computed as the product $A^T\cdot s$ followed by the addition of a constant value $h$ and a rounding operation. The public key is composed of the seed to generate $A$ and the sample $b$. The secret key is $s$.

The \textbf{encryption}, described in Algorithm~\ref{alg:Enc}, starts by regenerating the public matrix $A$. Then, it proceeds in the same way as the key generation to generate a new secret $s'$ and a new sample $b'$. Additionally, a sample $v'$ is computed as the product of the other part of the public key with the new secret $b^T\cdot s'$. The message $m$ is encoded in this vector $c_m$, which together with $b'$ constitutes the ciphertext. The \textbf{decryption}, described in Algorithm~\ref{alg:Dec}, computes a new sample $v$ in an analogous way to encryption by multiplying $b'^T\cdot s$. The message is recovered by decoding the difference between $v$ and a scaled version of the other part of the ciphertext $c_m$.

\begin{algorithm}[!ht]
    \caption{$\mathtt{Saber}{.}\mathtt{PKE}{.}\mathtt{KeyGen} ()$~\cite{saber}}
    \label{alg:KeyGenPKE}
    \begin{algorithmic}[1]

\State{$seed_{\pmb{A}} \leftarrow  \mathcal{U}(\{0,1\}^{256})$}
\State{$\pmb{A} = \mathtt{gen}(\text{seed}_{\pmb{A}}\text{)}  \in R^{l \times l}_q$}
\State{$r = \mathcal{U}(\{0,1\}^{256})$}
\State{$\pmb{s} = \beta_\mu ( R^{l \times 1}_q; r)$}
\State{$\pmb{b}=  ((\pmb{A}^T \pmb{s} + \pmb{h}) \bmod{q}) \gg (\epsilon_q - \epsilon_p) \in R^{l \times 1}_p$}\\
\Return{$(pk := (seed_{\pmb{A}}, \pmb{b}), \pmb{s})$}

    \end{algorithmic}
\end{algorithm}

\begin{algorithm}[!ht]
    \caption{$\mathtt{Saber}{.}\mathtt{PKE}{.}\mathtt{Enc} (pk = (\pmb{b}, seed_{\pmb{A}}), m \in R_2; r )$~\cite{saber}}
    \label{alg:Enc}
    \begin{algorithmic}[1]

\State{$\pmb{A} = \mathtt{gen}(\text{seed}_{\pmb{A}}\text{)} \in R^{l \times l}_q$}
\If {r  \text{is not specified}}
\State{$ r = \mathcal{U}(\{0,1\}^{256})$}
\EndIf
\State{$\pmb{s'} = \beta_\mu ( R^{l \times 1}_q ; r)  $}
\State{$\pmb{b}'= ( ( \pmb{A}  \pmb{s}' + \pmb{h}) \bmod{q})  \gg (\epsilon_q - \epsilon_p)  \in R^{l \times 1}_p$}\label{algline:sprime_enc_1}
\State{$ v'= \pmb{b}^T  (\pmb{s}' \bmod{p})  \in R_p$}\label{algline:sprime_enc_2}
\State{$c_m = (v'  + h_1 - 2^{\epsilon_p-1} m \bmod{p}) \gg (\epsilon_p - \epsilon_T) \in R_{T} $}\\
\Return{$c:=(c_m, \pmb{b'})$}

    \end{algorithmic}
\end{algorithm}

\begin{algorithm}[!ht]
    \caption{$\mathtt{Saber}{.}\mathtt{PKE}{.}\mathtt{Dec}(\pmb{s},c=(c_m, \pmb{b'}))$~\cite{saber}}
    \label{alg:Dec}
    \begin{algorithmic}[1]

\State{$v = \pmb{b} ^{\prime T} (\pmb{s} \bmod{p}) \in R_p $}
\State{$m' = ((v-2^{\epsilon_p-\epsilon_T}c_m + h_2) \bmod{p}) \gg (\epsilon_p-1) \in R_2$}\\
\Return{$m'$}

    \end{algorithmic}
\end{algorithm}

{\rd
\begin{algorithm}[!ht]
 \caption{ $\mathtt{Saber}{.}\mathtt{KEM}{.}\mathtt{KeyGen} ()$ }
 \label{alg:KeyGen}
 $(seed_{\pmb{A}}, \pmb{b}, \pmb{s})  = \mathtt{Saber}{.}\mathtt{PKE}{.}\mathtt{KeyGen} ()$ \\
 $pk = (seed_{\pmb{A}}, \pmb{b})$ \\
 $pkh = \mathcal{F}(pk) $ \\
 $z = \mathcal{U}(\{0,1\}^{256}$) \\
 \Return $(pk := (seed_{\pmb{A}}, \pmb{b}), sk :=  (z, pkh, pk, \pmb{s}))$
\end{algorithm}

}

\begin{algorithm}[!ht]
    \caption{$\mathtt{Saber}{.}\mathtt{KEM}{.}\mathtt{Encaps} (pk := (seed_{\pmb{A}}, \pmb{b}))$}
    \label{alg:Encaps}
    \begin{algorithmic}[1]

\State{$m  \leftarrow  \mathcal{U}(\{0,1\}^{256}) $}
\State{$(r, \hat{K}) = \mathcal{G}(\mathcal{H}(pk), m)$}
\State{$c = \mathtt{Saber}{.}\mathtt{PKE}{.}\mathtt{Enc} (pk, m, r )$}
\State{$K = \mathcal{H}(\mathcal{H}(c), \hat{K})$}\\
\Return{$(c, K)$}

    \end{algorithmic}
\end{algorithm}

\begin{algorithm}[!ht]
    \caption{$\mathtt{Saber}{.}\mathtt{KEM}{.}\mathtt{Decaps} (c, sk :=  (z, pkh, pk, \pmb{s}))$}
    \label{alg:Decaps}
    \begin{algorithmic}[1]

\State{$m'  =  \mathtt{Saber}{.}\mathtt{PKE}{.}\mathtt{Dec} (\pmb{s}, c )$}
\State{$(r',\hat{K}') = \mathcal{G}(\mathcal{H}(pk), m')$}
\State{$c' = \mathtt{Saber}{.}\mathtt{PKE}{.}\mathtt{Enc} (pk, m'; r')$}
\If{$c=c'$}
    \State{\textbf{return} $ K = \mathcal{H}(\mathcal{H}(c), \hat{K}')$}
\Else
    \State{\textbf{return} $ K = \mathcal{H}(\mathcal{H}(c), z)$}
\EndIf

    \end{algorithmic}
\end{algorithm}


The Chosen Ciphertext Attack-secure (CCA) version of Saber is achieved by applying the Fujisaki-Okamoto transform to the encryption scheme. Such construction requires two additional hash functions $\mathcal{G}$ and $\mathcal{H}$, which are SHA3-512 and SHA3-256, respectively. In its KEM setting, {\rd the \textbf{key-generation} algorithm utilizes the public-key key-generation algorithm as shown in {Algorithm~\ref{alg:KeyGenPKE}}. Additionally, it adds the hash of public-key, the public-key, and $256$-bit random number ($z$) for CCA security }. The message is randomly generated during \textbf{encapsulation} as shown in Algorithm~\ref{alg:Encaps}. The hash functions are used to generate randomness for the encryption as well as to generate the established session key. During \textbf{decapsulation}, the encryption is recalculated to check for potentially maliciously crafted ciphertexts. If the ciphertext match, the session key is computed in the same way using the hash functions as shown in Algorithm~\ref{alg:Decaps}.

\begin{figure}[!ht]
  \centering
  \includegraphics[width=0.48\textwidth]{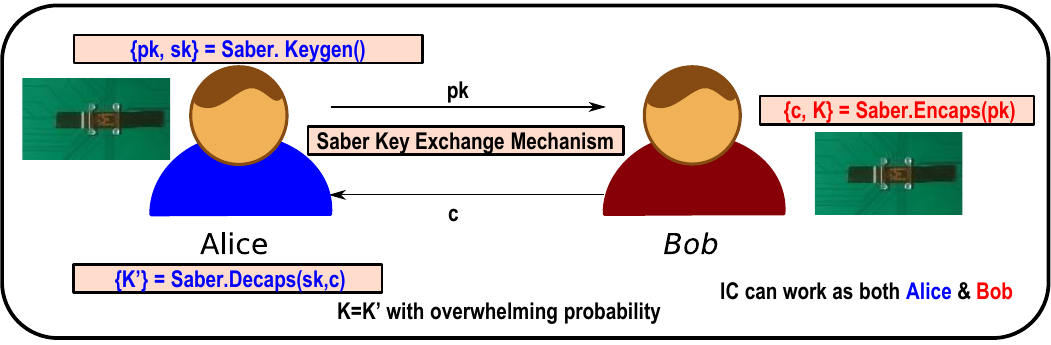}
  \caption{{\rd A sample KEM mechanism between 2 communicating parties. Note that our IC can act as both parties. Keygen(), Encaps() and Decaps() are presented in Algorithm 4, 5 \& 6 respectively.}}
  \label{KEM_keygen}
\end{figure}

{\rd A sample KEM is shown in Fig.~\ref{KEM_keygen} between 2 parties, namely Alice \& Bob. Alice creates the public key and secret key using algorithm ~\ref{alg:KeyGen}. She sends public key to Bob. Upon receiving that, bob calculates the encapsulation and sends the ciphertext of a message $m$ using algorithm~\ref{alg:Encaps}. Alice decapsulates using her secret key and received ciphertext c using algorithm ~\ref{alg:Decaps}. It should be noted that our IC can work as both Alice \& Bob here. }

\subsection{Features of Saber}
However, compared to the other lattice-based finalist schemes of NIST standardization procedure \textit{i.e.} Kyber~\cite{kyber_round_3} and NTRU~\cite{ntru_kem_hrss_nist}, Saber is relatively less studied scheme. Further, the design of Saber is also quite unique and unconventional compared to the other lattice-based schemes. Thus studying implementation aspects of Saber on dedicated hardware platforms ASIC is an appropriate and attractive field of investigation for the benefit of PQC. We describe some of the salient features of Saber below.
Here are 3 salient features of Saber as PQC-KEM scheme: 
\begin{itemize}
    \item {\emph{Power-of-2 moduli:} Polynomial multiplication is one of the most computationally expensive components of LBC. The \emph{de-facto} standard modulus for LBC is a prime that facilitates usage of fast quasi-linear number theoretic transform (NTT)~\cite{NTT_pollard} based polynomial multiplication. However, Saber uses an unorthodox choice of power-of-two modulus. It has been shown that Saber performance is at par with similar schemes which use NTT based polynomial multiplication on different platforms~\cite{DBLP:journals/tches/MeraKV20, DBLP:journals/tches/KarmakarMRV18, DBLP:journals/tches/ChungHKSSY21} by combining sub-quadratic polynomial multiplication algorithms Toom-Cook~\cite{toom,cook} and Karatsuba~\cite{karatsuba}. }
    \item {\emph{Flexibility:} Due to the use of module-lattices, the security of Saber can be easily upgraded or downgraded by appropriately increasing or decreasing the number of polynomials in the public matrix. This requires little to no change in other parameters of the scheme. Therefore, ASICs and software libraries designed for one level of security can be easily adapted to other security levels with small changes which results in great flexibility.}
    \item {\emph{Resistance to side-channel attacks:} Saber uses constant-time algorithms to prevent side-channel attacks like timing or simple-power analysis attacks. Generating noise inherently by rounding and using centered binomial distributions instead of more complex discrete Gaussian distributions~\cite{discrete_gaussian_dac, discrete_gaussian_TC} reduces the side-channel attack surface of Saber greatly. {\rd Also, it has been recently shown that masking, which is a provably secure countermeasure against powerful side-channel attacks, can be integrated much more efficiently on Saber than other lattice-based KEMs such as Kyber{~\cite{first_order_masked_saber},~\cite{DBLP:conf/scn/KunduDBKV22}}. This efficiency comes from the usage of power-of-2-moduli rather than prime moduli. In this work, we have followed constant-time implementations and avoided conditional branching on secret values similar to previous works on Saber{~\cite{saber}}. Therefore, our implementation is resistant to simple power analysis (SPA) or simple electromagnetic analysis (SEMA) attacks. Recently, physical countermeasures have been very popular in this context as they are architecture-agnostic \cite{singh_25.3_2019, ghosh202136,ghosh2021syn,ghosh2022digitalcicc}. Integrating these countermeasures (masking or physical countermeasures) along with attack detection mechanism (e.g. \cite{ghosh2022scadate, ghosh2022electromagnetic}) with this Saber architecture will help in more sophisticated DPA/CPA/CEMA security. These countermeasures can be integrated in the future versions of the ASIC for SCA-resilient implementation of Saber. }}
\end{itemize}
Though NIST has recommended Kyber for standardization for a few reasons, some of which are not strictly technical per se. 
Their final report~\cite{nist_third_round_report} (Sec.~4.3.4) mentions that the security, performance, bandwidth usage, etc. of Saber is at par with Kyber. Therefore Saber is perfectly suitable to be used as a PQC KEM in the future. Please note that another NIST finalist NTRU~\cite{ntru_kem_hrss_nist} is already present in many popular products like OpenSSH (version 9.0 onwards), GPL, WolfSSL, etc.

{\rd Furthermore, as Saber and Kyber are both based on module lattices and share a large number of individual blocks, we firmly believe that many of the {\it techniques} developed here for Saber can also be used for Kyber. For example, the polynomial sizes of Saber and Kyber are same and hence our strategies can be even used to realize a low-power and area instantiation of Kyber on ASIC. We should note that Kyber in its current form cannot be executed directly on our ASIC. To realize this, we need to make suitable changes such as the incorporation of a suitable modular reduction strategy, updating data-paths to suit the parameters of Kyber, deciding how the matrix A is generated, etc. These can be chosen according to the final goal of the ASIC design. However, this is an intriguing research question that needs more attention and due deliberation. Therefore, we leave this as future work.}
\subsection{PQC Hardware Implementations: State-of-the-Art}

In this section, we summarize the state-of-the-art of hardware implementations of post-quantum lattice-based KEMs. A more detailed analysis of performance, area and power figures is deferred to Section~\ref{sec:results}. The most frequent hardware implementations that can be found in the literature are FPGA-based implementations. The reason is that lattice-based cryptography has undergone a quick and recent development and the shortest design cycle of FPGA implementations is more adequate to develop a proof of concepts or to demonstrate algorithmic optimizations for existing schemes. Particularly, all most relevant lattice-based KEMs have FPGA implementations as can be seen for Frodo~\cite{DBLP:journals/jce/HoweMOR21}, Kyber~\cite{DBLP:conf/date/YamanMOS21}, NTRU~\cite{DBLP:journals/iacr/DangMG21}, NTRUPrime~\cite{DBLP:journals/iacr/PengMTYC21} and Saber~\cite{DBLP:conf/dac/MeraTKRV20}~\cite{DBLP:journals/tches/RoyB20}.

Despite the variety of schemes, the differences between distinct hardware implementations lie in the optimizations performed on the multiplier to suit better the parameters of the corresponding scheme. Focusing on Saber hardware implementations, we can distinguish three different approaches toward multiplication. \textit{First}, the Toom-Cook algorithm is used to accelerate the multiplication on hardware, together with an HW-SW co-design strategy to achieve a compact implementation on a heterogeneous ARM+FPGA System-on-Chip (SoC)~\cite{DBLP:conf/dac/MeraTKRV20}. \textit{Second}, high performance can be achieved with a fully parallel multiplier where the full polynomials are loaded and shifted after every multiplication~\cite{DBLP:journals/tches/RoyB20}. \textit{Third}, a combination of the Karatsuba algorithm and parallel multipliers has been proposed to achieve the high-performance operation of Saber while reducing the area requirements with respect to the fully parallel approach~\cite{DBLP:journals/iacr/ZhuZYZDCWL20}. Both the second and the third approach have been used to implement Saber on full custom hardware~\cite{DBLP:journals/tcasI/ZhuZYZDCWL21}~\cite{DBLP:journals/iacr/ImranABRP22}. 
{\rd Though \cite{DBLP:journals/iacr/ImranABRP22} proposes another hardware for Saber, as mentioned in the paper \cite{DBLP:journals/iacr/ImranABRP22}, due to a logic bug, the address offsets of 3 of the 4 distributed memories are incorrectly decoded and data is overwritten. This logical error results in a few flipped bits on the output of the chip when compared with the
expected results. Due to this issue, we refrain from comparing performance directly with this work~\cite{DBLP:journals/iacr/ImranABRP22} in the comparison table as the table includes only solid-state circuit literatures with functional ICs. }

\begin{figure*}[!ht]
  \centering
  \includegraphics[width=0.87\textwidth]{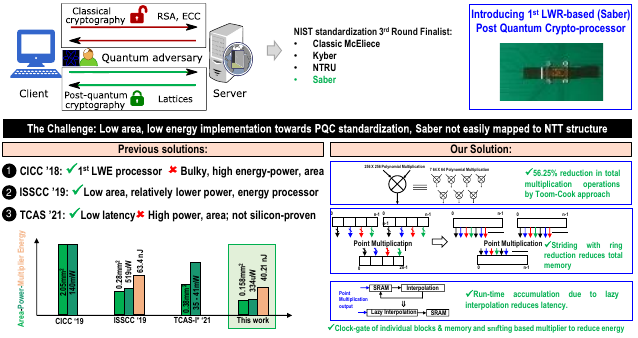}
  \caption{Classical PKC is broken by quantum computers. This work introduces the first silicon-verified M-LWR-based post-quantum crypto-accelerator. Fabricated IC uses a striding Toom-Cook-based polynomial multiplier with lazy interpolation to reduce latency and memory. Clock-gating of individual blocks and shift-based unit multiplier achieve the lowest power design. }
  \label{motivation_1}
\end{figure*}

\subsection{Motivation}\label{sec:intro_motiv}

{\rd KEM schemes are used to establish a common secret key between two or more communicating parties. A very well-known usage of KEM is inside transport layer security (TLS) protocol which is a newer version of the secure socket layer (SSL) protocol ~\cite{krawczyk2013security}. In TLS the \emph{handshake} protocol uses KEM to establish the common secret \emph{session} key. A very well-known application that uses TLS or by extension KEM scheme is HTTPS or Hypertext Transfer Protocol (HTTP) over SSL. In HTTPS the client (browser) first authenticates the server (web server) using the help of certificate authority and digital signature algorithms. The client also obtains the SSL/TLS certificate in this process. This certificate among other things contains the KEM scheme name and the public key of the server.  The client uses this KEM public key to initiate the handshake protocol which finishes with a common secret key on both sides of the client and server. This secret key is now used by a previously agreed symmetric-key algorithm like the advanced encryption standard (AES) to encrypt all the data communication between the client and the server. 

This secret key derived using the handshake protocol is also called \emph{master} secret. Whenever a client opens a connection with the server there might not be a full handshake involving the KEM protocol. Instead, the client and server can reuse the master secret key established during some previous connection. In this case, the secret key for the symmetric-key algorithm is derived using the master secret and a random value that the client and the server sent each other during their initial messages. This is called \emph{abbreviated handshake}. Multiple connections which share the same master secret constitute a session. Depending on the security settings of the client and server this master secret is kept alive for a long duration. Therefore removing the necessity of invoking the KEM algorithm for a long time. Hence, the KEM schemes are short-lived and invoked only a few times during the lifetime of the application.
}

However, the accelerator is there for the lifetime of the system. Due to the short-lived nature of KEMs in typical applications, the latency of KEM operations becomes less relevant compared to other metrics such as small area of the processor, low power, and low energy. High performance is arguably more important for symmetric key primitives that are constantly used once the communication has been established. A smart design of a KEM accelerator seeks to reduce the cost in the main system by achieving low area as well as to reduce the operational costs by achieving low power and low energy.

In this paper, we also aim at filling this gap in the state-of-the-art by bringing the first approach in silicon. Moreover, fundamental constraints such as low area and low energy when implementing key exchange operations makes the \textit{first approach}, i.e., Toom-Cook, more attractive for a dedicated IC. In Section~\ref{sec:results} we show different metrics to compare our design~\cite{ghosh2023334, ghosh2022334uw} to other Saber ASIC designs as well as to ASIC designs~\cite{DBLP:conf/cicc/SongTCZ18} that accelerate other lattice-based schemes such as NewHope~\cite{DBLP:conf/cicc/SongTCZ18} and Kyber~\cite{DBLP:journals/tches/BanerjeeUC19}.

\subsection{Contributions}
Figure~\ref{motivation_1} shows the key contributions of this work which we categorically describe below.
\begin{itemize}
    \item \textbf{First silicon-verified implementation of Saber:} 
    We have presented the first Saber core and till now one of the few \emph{fully functional} PQC ASIC. Therefore this work is a cornerstone in the development of PQC and in the transition from classical PKC to PQC.
    \item \textbf{Multiplier optimization with striding Toom-Cook and lazy interpolation to reduce energy and power of the accelerator:} Multiplication of Saber cannot be directly mapped with NTT structure, unlike Kyber. Different approaches have been taken to circumvent this problem. One approach is to use an alternative multiplication strategy. The second approach is to tweak the Saber algorithm so that it can be mapped to NTT structure~\cite{DBLP:journals/tches/ChungHKSSY21}. First approach is taken here. Several works are exploring classical Toom-Cook-based architecture. In this work, we have further optimized Toom-Cook multiplication with striding and lazy interpolation to make it memory efficient. It should be noted that this architecture helps us to reduce total SRAM used in the IC. Significant amount of power/energy consumption comes from the memory block which is leakage dominated. Reducing total memory size significantly helps to reduce final energy consumption. \\
    {\rd It should be noted that previously lazy interpolation is explored only in software ~\cite{DBLP:journals/tches/MeraKV20} in Toom-Cook multiplication context. ~\cite{DBLP:journals/tches/MeraKV20} focuses on a software Cortex-M4 based latency vs memory optimization problem for Toom-Cook multiplication. Additionally, our work merges a striding memory-access approach with lazy interpolation to provide a unified low-power, low-energy, and reduced memory ASIC for Saber with comparable latency for the multipliers.}
    \item \textbf{Re-configurable instruction-based multi-purpose architecture:} This architecture is re-configurable and can be controlled by micro-instruction provided from outside. Hence, the same architecture can be used to calculate encapsulation, decapsulation, and key generation removing the requirement for duplicate hardware. 
    \item \textbf{Lowest area \& power implementation among the PQC cores and accelerators:} This work provides 36\% power improvement from silicon-verified PQC core and 118X improvement with respect to state-of-the art Saber accelerator (not silicon verified) ~\cite{DBLP:journals/tcasI/ZhuZYZDCWL21}. Saber core takes 0.158 $mm^2$ of the area and 10.1875KB memory which is the lowest among PQC cores to date.  
\end{itemize}

\subsection{System Overview \& Paper Organization}
This IC has dedicated blocks for each micro-operations namely polynomial multiplication, binomial sampler, addpack, addround, cmov, verify etc. Polynomial multiplication is the most computationally complex, hence most area \& power hungry block. This is chosen for algorithmic \& architectural optimization in order to achieve low area and low energy.  
Algorithmic level optimizations of the multiplier are discussed in Section~\ref{sec:polymul}. Section~\ref{sec:hw} discusses hardware optimizations along with a brief discussion about all the circuit components. Silicon results are presented in Section~\ref{sec:results} before we conclude the work in Section~\ref{sec:conclusions}.


\section{Multiplier Optimizations}\label{sec:polymul}

The core operation of Saber as a M-LWR scheme is the matrix-vector multiplication between the public matrix $A$ and the secret vector $s$. Since the dimension of the module lattice is $l=3$, the public matrix $A$ is composed by $3\times 3$ polynomials with $256$ coefficients each. Therefore, the operation that needs to be optimized is actually the polynomial multiplication, which is defined in the algebraic ring $R_{q} = \mathbb{Z}_q[x]/(x^{n}+1)$. Mathematically, a multiplication {\rd in} $R_q$ is a standard polynomial multiplication followed by the modular reduction by $(x^{n}+1)$. The rule of thumb to compute the modular reduction is to {\rd equalize} the modulus to zero $(x^{n}+1)$ and, therefore, to apply the change of variable $x^n=-1$ to the product of two polynomials. In other words, the multiplication in $R_q$ is equivalent to a negatively wrapped polynomial multiplication. Later n this section, we explain how to implement the multiplier to exploit this structure. Next, we discuss the algorithmic choices for polynomial multiplication.

There are two approaches to implement polynomial multiplication on hardware. First, one can use the straightforward algorithm with quadratic complexity referred to as \textbf{schoolbook}~\cite{DBLP:journals/tches/RoyB20}. While this approach leads to poor performance figures on software, on hardware we can take advantage of its simpler structure to parallelize the arithmetic operations. The degree of parallelization can be increased to trade off area for higher performance and research shows that the highest performance for polynomial multiplication can be achieved with the fully parallel multiplier~\cite{DBLP:journals/iacr/DangMG21}.

Second, one can optimize the polynomial multiplication algorithm to reduce the computational complexity. Then, the implementation of the chosen algorithm can still be optimized at platform level to take advantage of the available resources. This approach is the most popular for software implementations~\cite{DBLP:conf/dac/MeraTKRV20}~\cite{DBLP:journals/tches/ChungHKSSY21} but it can also be followed with successful results on hardware~\cite{DBLP:conf/dac/MeraTKRV20}. 

The polynomial multiplication module is implemented following the second approach. Moreover, we use Toom-Cook $4$-way to reduce the complexity of the polynomial multiplication and parallelize it as in~\cite{DBLP:conf/dac/MeraTKRV20}. Thus, for Saber parameters a single $256\times 256$ polynomial multiplication is broken down into seven $64\times 64$ polynomial multiplications, which are parallelized at hardware level.
We propose additional optimizations to the multiplication, a \textbf{strided algorithm} and \textbf{lazy interpolation}, both of which are explained next in this section. There are three benefits of this approach: 
\begin{itemize}
    \item Using Toom-Cook multiplication reduces total number of multiplication from $256\times 256$  into seven $64\times 64$ polynomial multiplications hence saving 56.25\% of total multiplication operation. 
    \item Striding Toom-Cook reduces the memory requirement, hence helps in reducing energy consumption and total area.
    \item Lazy interpolation helps in reducing number of iteration and helps in latency improvement. 
\end{itemize}

\subsection{Classical Toom-Cook vs Striding Toom-Cook}
\begin{algorithm}[!ht]
    \caption{\color{black}Evaluation of \color{violet}classical \color{black}and \color{blue} striding \color{black} Toom-Cook 4-way with vertical scanning}
    \label{alg:tc_eval_merge}
    \begin{algorithmic}[1]
        \Require{Two arrays $A$ and $B$ with the $n=256$ coefficients of the polynomials}
        \Ensure{Seven arrays $w_1$ to $w_7$ with \color{violet}$127$\color{black}{ }\slash{ }\color{blue}$64${ }\color{black} coefficients of the intermediate products each}

        \For{$j=0\ to\ 63$}
            \State{$r_0=A_{0}[j]$};
            \State{\color{violet}$r_1=A_{64}[j]$;\;\;\;\;\;\;\;\color{blue}$r_1=A_{1}[j]$};
            \State{\color{violet}$r_2=A_{128}[j]$;\;\;\;\;\;\color{blue}$r_2=A_{2}[j]$};
            \State{\color{violet}$r_3=A_{192}[j]$;\;\;\;\;\;\color{blue}$r_3=A_{3}[j]$};
            \State{$r_4=r_0+r_2$;\;\;\;\;\;$r_5=r_1+r_3$;}
            \State{$r_6=r_4+r_5$;\;\;\;\;\;$r_7=r_4-r_5$;}
            \State{$aws_3[j]=r_6$;\;\;\;\;\;$aws_4[j]=r_7$;}
            \State{$r_4=2*(r_0*4+r_2)$;}
            \State{$r_5=r_1*4+r_3$;}
            \State{$r_6=r_4+r_5$;\;\;\;\;\;$r_7=r_4-r_5$;}
            \State{$aws_5[j]=r_6$;\;\;\;\;\;$aws_6[j]=r_7$;}
            \State{$r_4=8*r_3+4*r_2+2*r_1+r_0$;}
            \State{$aws_2[j]=r_4$;}
            \State{$aws_7[j]=r_0$;\;\;\;\;\;$aws_1[j]=r_3$;}
        \EndFor
        \State{Repeat the above steps to generate the weighted polynomials $bws_1$ to $bws_7$}
        \For{$i=1\ to\ 7$}
            \State{$w_i=aws_i*bws_i$};
        \EndFor\\
        \Return{$w_1$ to $w_7$}

    \end{algorithmic}
\end{algorithm}

Toom-Cook $k$-way uses a divide-and-conquer approach to break down a single multiplication of polynomials with $n$ coefficients each into $2k-1$ multiplications where the new operands have $n/k$ coefficients each instead. The procedure to generate the intermediate operands is called evaluation. The final result can be retrieved applying the inverse transformation to evaluation, namely interpolation. Mathematically, the inputs to the multiplication are lifted from $R\left [ x \right ]$ to the isomorphic ring $R\left [ x \right ]\left [ y \right ]/(x^k-y)$. Then, the operands can be expressed as $a(x) = (a_0+a_1x+\cdots +a_{k-1}x^{k-1})+\cdots+(a_{n-k}+\cdots+a_{n-1}x^{k-1})y^{r-1}$ where $r=n/k$. Particularizing $n=256$ and $k=4$, we can derive Algorithm~\ref{alg:tc_eval_merge} with violet text to perform Toom-Cook evaluation on points $x=\left \{ 0,1,-1,1/2, -1/2 , 2, \infty  \right \}$. This algorithm incorporates the vertical scanning method proposed in~\cite{DBLP:journals/tches/KarmakarMRV18} to reduce the overhead introduced by memory operations. On hardware this method allows the parallelization of the evaluation to build all seven intermediate polynomials simultaneously.

However, there are two factors that penalize the performance and memory of Toom-Cook when implemented this way. First, the memory access pattern does not follow any spatial locality. The sequence of the coefficients indexes accessed in classical Toom-Cook has offsets of $64$. When following a HW/SW co-design approach this can be solved by transferring the polynomials with the appropriate layout~\cite{DBLP:conf/dac/MeraTKRV20}, but this is not efficient when the whole scheme is accelerated in hardware. Second, this Toom-Cook algorithm requires size doubling in the intermediate polynomials.

We address the two inefficiencies of Toom-Cook by using a less known variant of Toom-Cook~\cite{Bernstein01multidigitmultiplication} referred to as striding Toom-Cook. In this variant, a different ring isomorphism is used. The inputs are lifted from $R\left [ x \right ]$ to $R\left [ y \right ]\left [ x \right ]/(x^k-y)$ instead of $R\left [ x \right ]\left [ y \right ]/(x^k-y)$. Since the ring of the original multiplication is $R\left [ x \right ]/(x^n-y)$ and $y=x^k$, the base ring for the intermediate multiplications becomes $R\left [ y \right ]/(x^r+1)$ which is also a negacyclic polynomial ring. Therefore, the modular reduction corresponding to the ring multiplication can be deferred to the point multiplication of Toom-Cook. Mathematically, the operands of this Toom-Cook variant can be written as $a(x) = (a_0+a_ky+\cdots+a_{(r-1)k}y^{r-1})+\cdots+(a_{k-1}+a_{2k-1}y+\cdots+a_{(r-1)k+k-1}y^{r-1})x^{k-1}$ where $y=x^k$ and $n=k\cdot r$. Again, we take into account the parameters of our multiplication, $n=256$, $k=4$, and we evaluate the polynomials in the same points as for classical Toom-Cook, $x=\left \{ 0,1,-1,1/2, -1/2 , 2, \infty  \right \}$. Thus, Algorithm~\ref{alg:tc_eval_merge} with blue text can be derived as evaluation of striding Toom-Cook. If we compare both versions, i.e., Algorithm~\ref{alg:tc_eval_merge} with violet or blue text, the only difference between classical Toom-Cook and striding Toom-Cook evaluation lies in the load operations that happen at the beginning of every iteration.
We can observe that the striding version reads coefficients with consecutive indexes which allows a more efficient hardware implementation.

\begin{algorithm}[!ht]
    \caption{\color{black}Interpolation of \color{violet}classical \color{black}and \color{blue} striding \color{black} Toom-Cook 4-way}
    \label{alg:tc_interpol_merge}
    \begin{algorithmic}[1]
        \Require{Seven arrays $w_1$ to $w_7$ with \color{violet}$127$\color{black}{ }\slash{ }\color{blue}$64${ }\color{black} coefficients of the intermediate products each}
        \Ensure{An array with the coefficients of $C=A(x)*B(x)$}

        \color{violet}\State{$C\leftarrow 0$}\color{black}
        \For{$i=0\ to\ $\color{violet}$126$\color{black}{ }\slash{ }\color{blue}$63${ }\color{black}}
            \color{blue}\State{$r_7=r_0$; \ \ $r_8=r_1$; \ \ $r_9=r_2$;}\color{black}
            \State{$r_1=w_2[i]$;\ \ \ \ $r_4=w_5[i]$;\ \ \ \
            $r_5=w_6[i]$;\ \ \ 
            $r_0=w_1[i]$;}
            \State{$r_2=w_3[i]$;\ \ \ \ $r_3=w_4[i]$;\ \ \ \
            $r_6=w_7[i]$;}
            
            \State{$r_1=r_1+r_4$;\ \ \ \ 
            $r_5=r_5-r_4$;\ \ \ \ 
            $r_3=(r_3-r_2)/2$;}
            \State{$r_4=r_4-r_0$;\ \ \ \ 
            $r_8=64\cdot r_6$;\ \ \ \  
            $r_4=2\cdot r_4+r_5$;}
            \State{$r_4=r_4-r_8$;\ \ \ \ $r_2=r_2+r_3$;\ \ \ \ $r_1=r_1-65\cdot r_2$;}
            \State{$r_2=r_2-r_6$;\ \ \ \ $r_2=r_2-r_0$;\ \ \ \  $r_1=r_1+45\cdot r_2$;}
            \State{$r_4= (r_4-8\cdot r_2)/_{24}$;\ \ \ \ $r_5=r_5+r_1$;}
            \State{$r_1=(r_1+16\cdot r_3)/_{18}$;\ \ \ \ $r_3=-(r_3+r_1)$;}
            \State{$r_5=(30\cdot r_1-r_5)/_{60}$;\ \ \ \ $r_2=r_2-r_4$;}
            \State{$r_1=r_1-r_5$;}
            
            \color{violet}\State{$C[i] += r_6$; \ \ $C[i+64] += r_5$;}
            \State{$C[i+128] += r_4$; \ \ $C[i+192] += r_3$;}
            \State{$C[i+256] += r_2$; \ \ $C[i+320] += r_1$;}
            \State{$C[i+384] += r_0$;}\color{black}

            \color{blue}\State{$C[4i+3] = r_3$;}
            \If{i == 0}
                \State{$C[4i] = r_6$; \ \ $C[4i+1] = r_5$; \ \ $C[4i+2] = r_4$;}
            \Else
                \State{$C[4i] = (r_6 +r_9)$; \ \ $C[4i+1] = (r_5+r_8)$;}
                \State{$C[4i+2] = (r_4+r_7)$;}\color{black}
            \EndIf
        \EndFor
        \color{blue}\State{$C[0] -= r_2$; \ \ $C[1] -= r_1$; \ \ $C[2] -= r_0$;}\color{black}
        \color{violet}\State{$C\leftarrow C(x)\mod (x^n+1)$}\color{black}
        \\
        \Return{$C$}

    \end{algorithmic}
\end{algorithm}

\begin{figure*}[!ht]
  \centering
  \includegraphics[scale=.5]{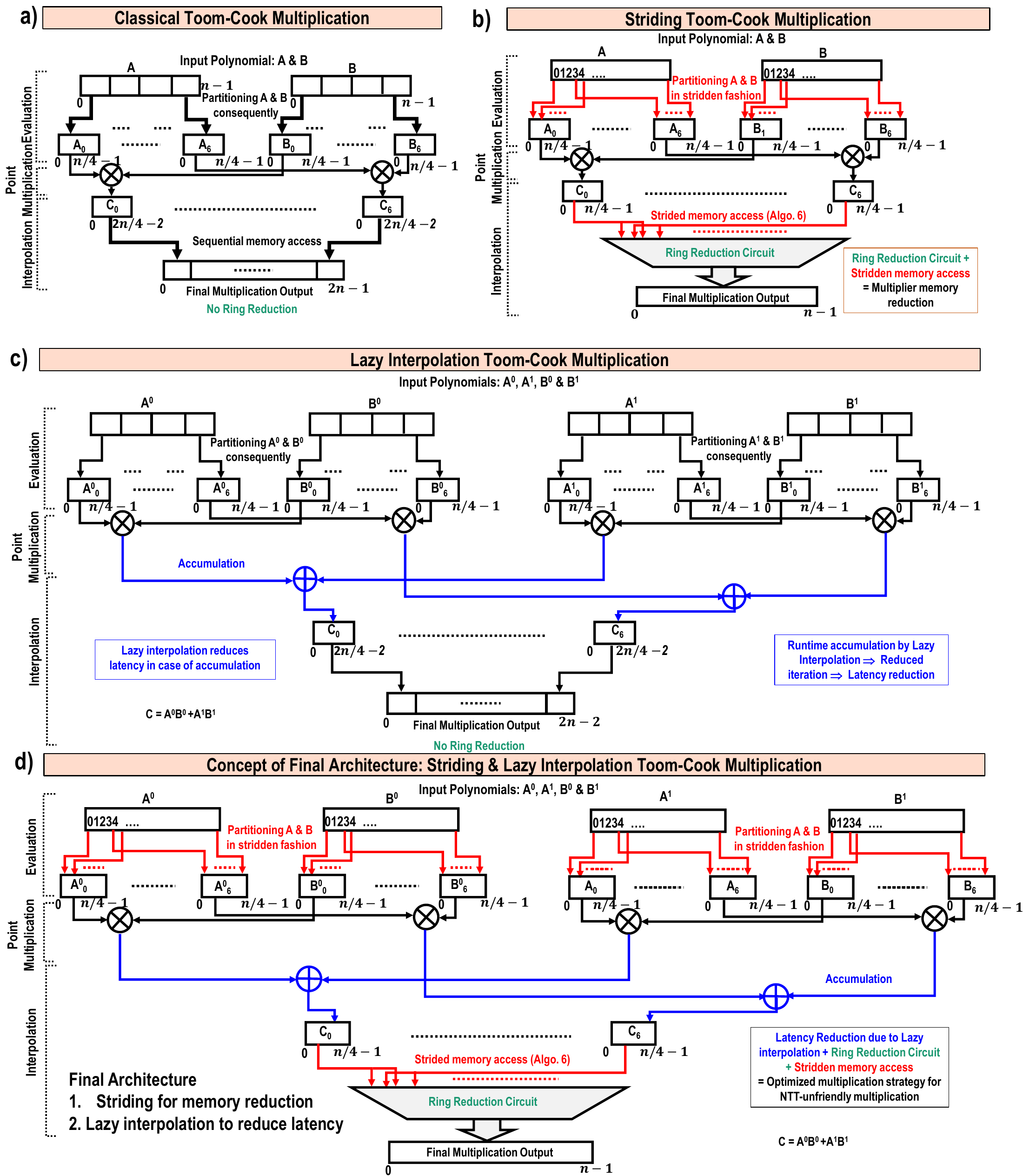}
\caption{a) Classical Toom-Cook architecture. b) Striding Toom Cook architecture that reduces memory footprint using a strided memory access. c) Lazy interpolation in classical Toom-Cook that reduces latency in interpolation with on-the-fly accumulation. d) Striding Toom-Cook with lazy interpolation, final implemented architecture to minimize both memory footprint and latency. }
  \label{fig:tc_vs_strd}
\end{figure*}

\begin{figure}[!ht]
  \centering
  \includegraphics[width=0.95\columnwidth]{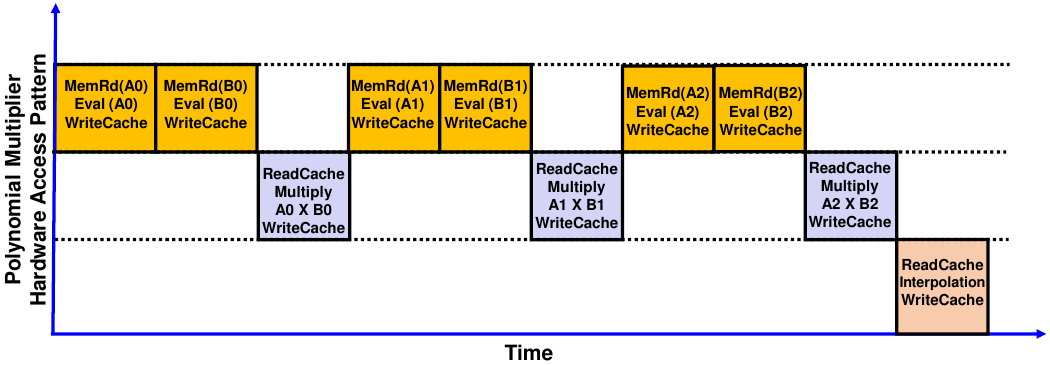}
\caption{{\rd Scheduling of polynomial multiplication.}}
  \label{fig:scheduling}
\end{figure}

The interpolation stage within the Toom-Cook multiplication is the inverse operation of the evaluation. Toom-Cook $k$-way works by evaluating the two operands in $2k-1$ different points to reduce the complexity of the polynomial multiplication. Once the result is obtained in the point-value domain, we need to interpolate these points to recover the coefficients of the polynomial. The evaluation can be defined as a matrix-vector multiplication where each row of the matrix is formed by the powers of the chosen point $((x_0)^0, (x_0)^1,\cdots, (x_0)^{2k-2})$ and the vector is formed by the coefficients of the polynomial. Therefore, the interpolation matrix is the inverse of the evaluation matrix. Moreover, it has been shown~\cite{DBLP:conf/issac/BodratoZ07} that the sequence of operations determined by the Gaussian elimination method to invert the matrix yields the optimal sequence of operations performed on the intermediate polynomials to interpolate the result. Thus, Algorithm~\ref{alg:tc_interpol_merge} with violet text is derived as the classical Toom-Cook interpolation.

Since the evaluation points are the same for classical and striding Toom-Cook, the evaluation and interpolation matrices are also the same. Consequently, the sequence of operations to compute the interpolation, i.e., lines~4-29 in Algorithm~\ref{alg:tc_interpol_merge}, is also the same for both versions. However, the access pattern to the coefficients is different. As explained when discussing evaluation, the striding version performs the ring reduction implicitly, prevents size doubling during multiplication and allows sequential access to the coefficients. Algorithm~\ref{alg:tc_interpol_merge} with blue text shows the striding interpolation. If we compare classical and striding versions of interpolation we can observe four differences. First, in classical Toom-Cook the array where the result is stored is initialized to $0$, see line~1 of Algorithm~\ref{alg:tc_interpol_merge}. This is because the ring reduction must be performed explicitly due to the size-doubling property of classical Toom-Cook. Second, thanks to the lack of size doubling the interpolation of striding Toom-Cook iterates over $64$ coefficients instead of $127$, see line~2 of Algorithm~\ref{alg:tc_interpol_merge}. This has an impact on memory compactness as well as on performance, which is particularly relevant in hardware where the latency of interpolation is comparable to the intermediate multiplications~\cite{DBLP:conf/dac/MeraTKRV20}. Third, the coefficients of the result are accumulated with $64$ coefficient offsets in the classical version while they are consecutive coefficients in the striding version, see lines~30-33 in Algorithm~\ref{alg:tc_interpol_merge} in contrast to lines~34-39. Accessing consecutive coefficients is more beneficial since it enables a higher throughput in memory operations. Lastly, in the classical version the ring reduction must be performed explicitly after the multiplication due to the size doubling. In the striding version, the output polynomial already belongs to the polynomial ring since the ring reduction happens inherently to the algorithm.
Figure~\ref{fig:tc_vs_strd} compares the full multiplication using classical Toom-Cook (a) and striding Took-Cook (b). Both methods are described visually and the main differences are highlighted in the figure.

\subsection{Toom-Cook and Lazy Interpolation}

Lazy interpolation and its application to software implementations of module lattice-based cryptography has been formalized in~\cite{DBLP:journals/tches/MeraKV20}. Lazy interpolation takes advantage of the fact that Toom-Cook evaluation and interpolation are linear transformations. Therefore, operations performed in the Toom-Cook domain are equivalent to operations performed after interpolation. In the setting of Saber, we exploit the matrix-vector structure to compute the entire row-column product in the Toom-Cook domain, thus deferring the interpolation to the end. This way we trade off all but the last interpolation operations in each row-column product at the expense of extra storage in the Toom-Cook domain. A visual representation of this technique is shown in Figure~\ref{fig:tc_vs_strd}(c). On the one hand, saving up interpolation operations becomes particularly relevant in hardware implementations where the latency of interpolation can be comparable to the latency of the intermediate multiplication of Toom-Cook. On the other hand, since the intermediate multiplications within Toom-Cook are parallelized anyway we do not incur in any memory penalization for utilizing lazy interpolation in our design. Additionally, we use lazy interpolation in combination with striding Toom-Cook to achieve an efficient hardware implementation of polynomial multiplication based on Toom-Cook.

A simplified schematic view of our proposed polynomial multiplication combining striding Toom-Cook with lazy interpolation is shown in Figure~\ref{fig:tc_vs_strd}(d). For simplicity, the impact of each optimization is shown on a row-column multiplication of dimension only $2$. We highlight with violet the changes due to the striding version of Toom-Cook and with teal the changes due to the use of lazy interpolation.

{\rd A simplified scheduling algorithm is presented in Fig.~\ref{fig:scheduling}. Once the polynomial multiplier is enabled, it accesses system memory to access the polynomials and evaluate it. The evaluated polynomial is stored in Cache memory. After that, point multiplication is activated to access the evaluated coefficients and multiplication outputs are again stored in the Cache. Finally, at interpolation stage, cache memory is accessed and interpolated values are stored back in system memory. It should be noted that due to run-time lazy interpolation, this stage happens once after 3 evaluations and point multiplications as Saber needs $3\times 3$ A matrix\cite{DBLP:journals/tches/MeraKV20}.  }


\section{System Architecture \& Optimizations}\label{sec:hw}

\begin{figure*}[!ht]
  \centering
  \includegraphics[width=\textwidth]{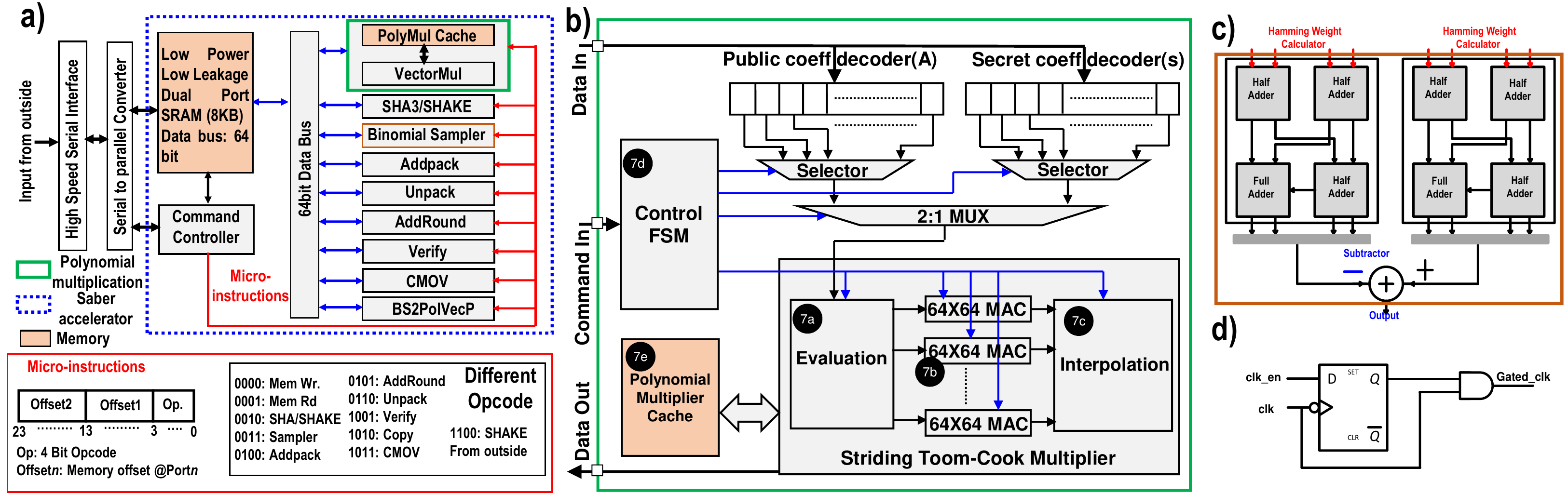}
  \caption{{a) The full system architecture of Saber core including opcode structure. b) Striding Toom-Cook multiplier architecture. c)~Hamming weight-based binomial sampler. d) Clock-gating circuit used in each block to minimize power consumption.}}
  \label{full_architecture_saber}
\end{figure*}

Figure~\ref{full_architecture_saber}(a) shows the fully implemented architecture for Saber. The high-speed serial interface is used as the chip interface from the outside. The serial-to-parallel converter changes the serial input to 64-bit parallel data as memory is implemented with the 64-bit data bus. Different blocks are controlled by the command controller, which takes the input from outside and breaks it down to the command format. Based on the command, different blocks activate themselves independently. The most area and power-hungry block is the polynomial multiplier block, highlighted with green boundaries in Figure~\ref{full_architecture_saber}(a). Multiplier optimization is discussed in the next subsection. The full multiplier architecture with striding Toom-Cook and lazy interpolation is presented in Figure~\ref{full_architecture_saber}(b) SHA3/SHAKE is implemented using the standard Keccak core provided by Keccak team~\cite{keccak_hw}. 
Binomial sampler is another important block towards any PQC core security. This is implemented in simple combinatorial xor gates to reduce power overhead as shown in Figure~\ref{full_architecture_saber}(c). The total memory requirement at the system level is 8KB. The system memory block is implemented using TSMC 65nm low power low leakage SRAM cell. It has 1K addresses with 64-bit words which is implemented continuously. Standard clock gating as shown in Figure~\ref{full_architecture_saber}(d) has been introduced to reduce leakage power further in each block, including system memory and polynomial multiplication caches.

A 24-bit micro-instruction format is used as a command to control different blocks. 4-bit opcode is used for enabling different blocks. Address offset1 and address offset2 are used as the input and output of a block respectively. However, polynomial multiplication needs two operands as inputs. Both input offset addresses are taken from both the offsets and the output offset starts from offset2 in this case.

\subsection{Multiplier}\label{hw:mult}

The polynomial multiplier is the most power and area-hungry block even after multiple optimizations as shown in the literature~\cite{DBLP:journals/tches/RoyB20}~\cite{DBLP:journals/tches/BanerjeeUC19}. We have observed the same tendency from baseline implementations which has motivated us to optimize the polynomial multiplication at algorithmic, architecture and circuit level. Multiple works earlier~\cite{DBLP:conf/cicc/SongTCZ18}~\cite{DBLP:journals/tcasI/ZhuZYZDCWL21} have shown low latency implementations. However, key exchange mechanism is used only at the beginning of the secret communication. Hence, when co-processor cores run at hundreds of MHz, i.e., latency in the order of milliseconds, higher latency is less important as much as achieving low area or lower energy consumption.

Multiplier architecture is depicted in Figure~\ref{full_architecture_saber}(b). Two decoders are there for public matrix $A$ and secret $s$. Decoders are required as data comes in packed format from the data memory. The decoder selects 13-bit or 4-bit of data as one coefficient of the public or secret polynomial, respectively. The data is fed to evaluation datapath for pre\-processing. Processed data is used by 64$\times$64 MAC units to perform the intermediate products of Toom-Cook algorithm. The coefficients of such intermediate products are cached until interpolation happens. Different datapaths are discussed in subsequent subsections. Evaluation, MAC, and interpolation are implemented according to the striding Toom\-Cook multiplication architecture. A small memory cache is used for internal data storage during polynomial multiplication operations. A control FSM controls all the sub-operations within multiplication in a timely fashion.
\begin{figure*}[!ht]
  \centering
  \includegraphics[width=0.9\textwidth]{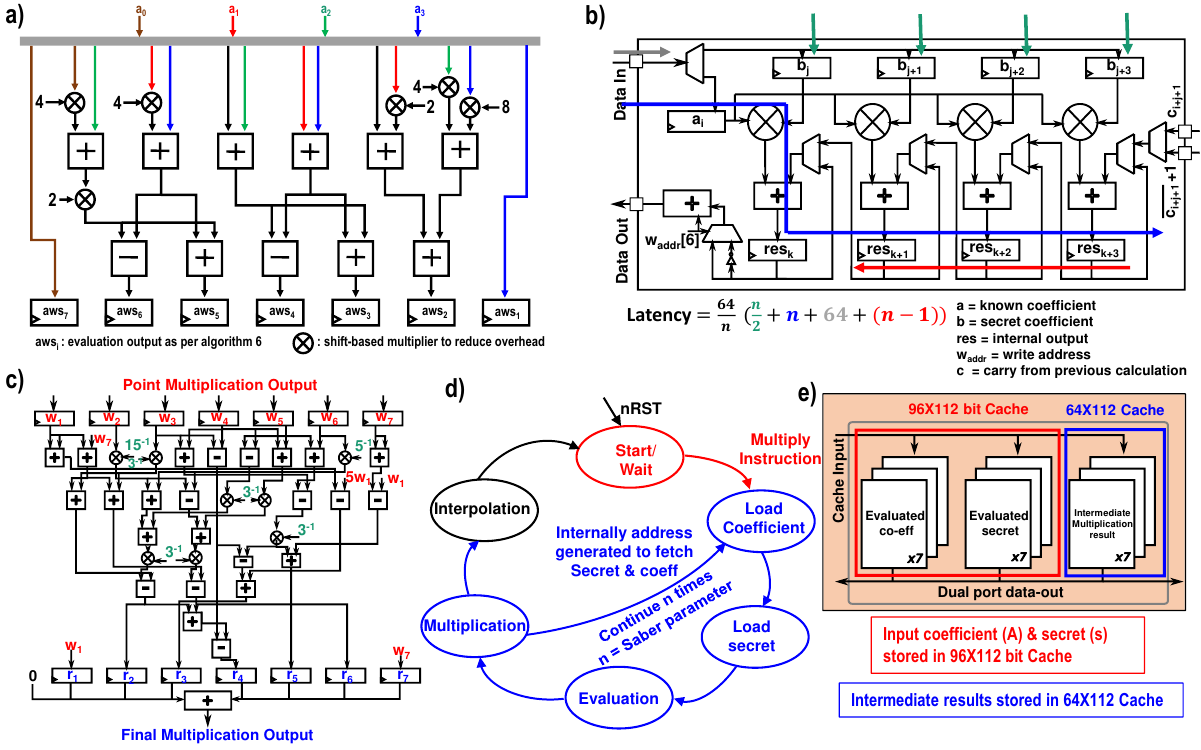}
  \caption{a) Evaluation datapath of striding Toom-Cook multiplier. b) Unit multiplication of striding Toom-Cook multiplier. Latency is optimized to be comparable with \cite{banerjee20192}. c) Interpolation datapath of striding Toom-Cook. d) Finite state machine to control full vector multiplication. e) Cache structure of the striding Toom-Cook multiplier.}
  \label{multiplication_blocks}
\end{figure*}

\subsubsection{Evaluation Datapath}


Evaluation is the first step of striding Toom-Cook multiplication. Public matrix and secret key are fed to evaluation datapath, which is implemented using Algorithm~\ref{alg:tc_eval_merge} as shown in Figure~\ref{multiplication_blocks}(a). Since the four coefficients used in every iteration of evaluation are consecutive, we can take advantage of 64-bit reads to fetch all four in a single clock cycle. The arithmetic operations performed during evaluation are additions of depth 2 in the worst case and multiplications by powers of 2 that can be implemented as simple bit shifts. Therefore, no additional pipelining is required to achieve perfect throughput as well as to reduce power and area. The latency of a single polynomial evaluation is only $64$ clock cycles for generating the $7$ intermediate polynomials in parallel.

\subsubsection{Multiply and Accumulation Units}

Saber performs 256$\times$256 polynomial multiplications. Striding Toom-Cook splits each 256$\times$256 multiplication into 7 64$\times$64 polynomial multiplications. Each of these 7 multiplications is performed in parallel by a Multiply and Accumulation Unit as presented in Figure~\ref{multiplication_blocks}(b). Next we explain how to choose the number of parallel multipliers in every MAC Unit.

Our design is optimized to equate latency with respect to state-of-the-art NTT implementations~\cite{DBLP:journals/tches/BanerjeeUC19} despite being lower energy and area implementation. If we assume $n$ number of parallel multipliers in the MAC architecture, hence, all the operations will be carried out $\frac{64}{n}$ times. Now, data is fetched from dual port memory. Hence, coefficients from $b$, as shown in Figure~\ref{multiplication_blocks}(b), are fetched in $\frac{n}{2}$ clock cycles. $n$ clock cycles are required to fill up the full datapath. After filling up the datapath, 64 cycles are required to perform the calculation as all 64 coefficients are multiplied. $n-1$ clock cycles are required to flush out the datapath. Total latency of the point multiplication is $\frac{64}{n}\times (\frac{n}{2}+n+64+(n-1))$. As $n$ = 4, 1168 clock cycles will be required which is comparable with respect to PQC core published in~\cite{DBLP:journals/tches/BanerjeeUC19} despite not being an NTT-based architecture.


\subsubsection{Interpolation Datapath}


The interpolation datapath can be derived in an equivalent manner as the evaluation datapath by directly mapping the operations in Algorithm~\ref{alg:tc_interpol_merge} to hardware. However, this would lead to a long critical path that would result in higher area and power consumption due to additional pipelining. Instead, we reorder the operations in lines~11-29 of Algorithm~\ref{alg:tc_interpol_merge} to reduce the depth of the circuit. In Section~\ref{sec:polymul} it is explained that this sequence of operations is obtained from applying Gauss elimination to invert the evaluation matrix. We find a different sequence of operations that produces the same result where the depth is optimized instead of the number of operations. This new sequence of operations is equivalent to the previous one and is matched directly to the hardware shown in Figure~\ref{multiplication_blocks}(c). This datapath is pipelined in three stages to match the frequency requirements of the rest of the circuit.

\subsubsection{Finite State Machine for Multiplication} 


Figure~\ref{multiplication_blocks}(d) sketches the finite state machine that controls multiplication. Active low reset is used to enable the FSM. After reset the FSM waits in start/wait state. As soon as a multiply instruction comes, it goes to next state. Next states are used to load the coefficients of the public polynomial and the secret. Then, evaluation followed by unit multiplication is performed. MAC units get enabled when the FSM enters the Multiplication state. This step continues in the loop for $n$ times. This $n$ is a cryptographic parameter, e.g., $n=3$ for Saber. The final state is interpolation, which is performed only once for each row-vector multiplication according to the lazy interpolation optimization. After that, the FSM waits for the next row-column multiply instruction.





\begin{figure}[!ht]
  \centering
  \includegraphics[scale=1]{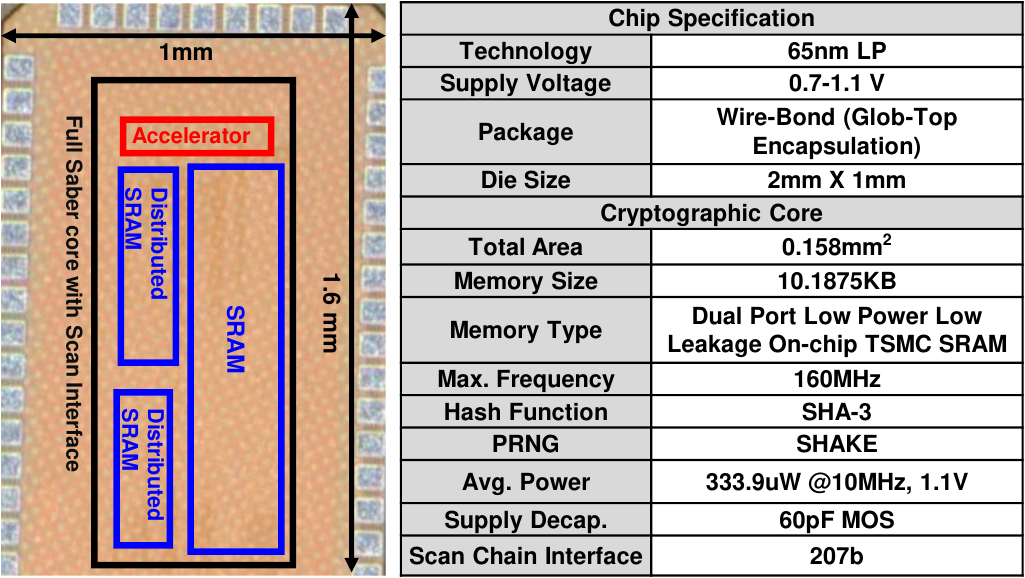}
  \caption{IC micrograph and specification.}
  \label{chip_spec}
\end{figure}

\begin{figure*}[!ht]
  \centering
  \includegraphics[width=0.95\textwidth]{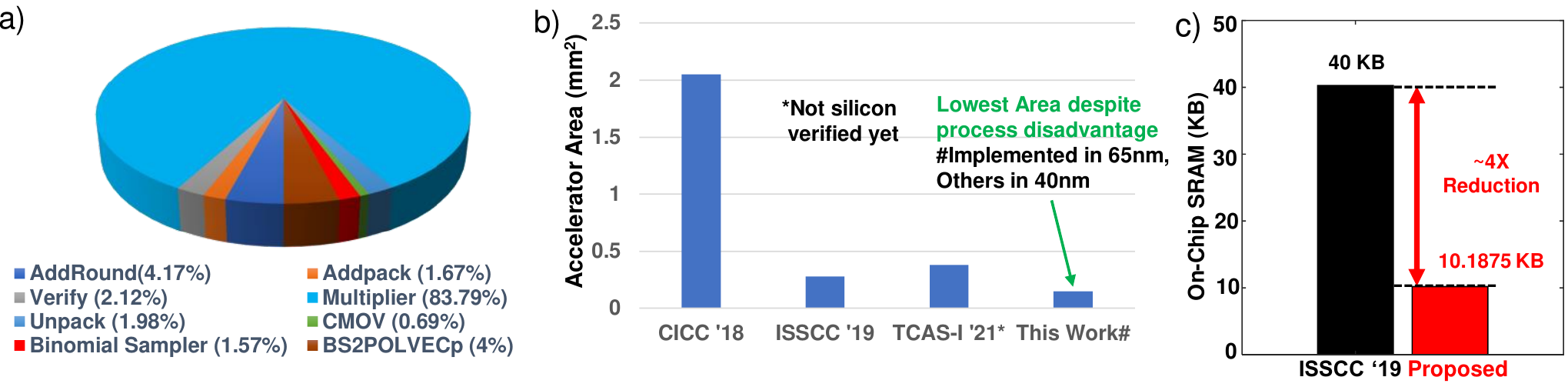}
  \caption{\red (a) Area requirement of different blocks. (b) Area comparison with state-of-the-art. Our design achieves the lowest area despite process disadvantages. (c) Memory footprint improvement with respect to the state-of-the-art. 
  }
  \label{silicon_area}
\end{figure*}
\begin{figure}[!ht]
  \centering
  \includegraphics[width=0.45\textwidth]{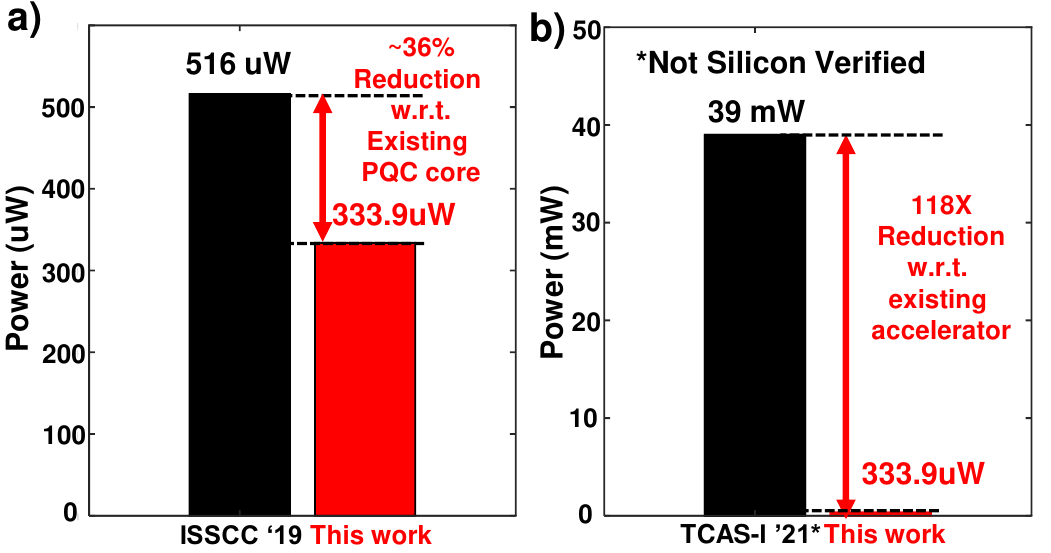}
  \caption{Power comparison with respect to state-of-the-art.}
  \label{power_comparison}
\end{figure}
\subsubsection{Memory Components for Multiplication}
Memory components are implemented using separate memory as cache. Key idea is to avoid stall cycles in the general operation. The cache is implemented using low power low leakage SRAM cells. The coefficients of the evaluated polynomial and secret are stored in a 96 $\times$ 112 cache as shown in the red border of Figure~\ref{multiplication_blocks}(e). Intermediate results are stored in a 64$\times$112 cache implemented with the same type of SRAM cells. This multiplication memory is clock gated when data memory is enabled and multiplication FSM is at wait state. This helps us to reduce energy when multiplication is not operational. Clock gating circuit is shown in Figure~\ref{full_architecture_saber}(d).

\subsection{Sampler}
Secret polynomials in Saber follow a discrete binomial distribution. The sampler module creates this distribution from uniformly distributed data. First, data is sampled from the PRNG. This data is taken nibble(4-bit) wise and hamming weight is calculated from that. Difference between hamming weight is used as sampler output. This sampling is combinatorial. Hamming distance is calculated using a half adder and full adder as shown in Figure~\ref{full_architecture_saber}(c). 

\subsection{Memory Components}

Memory components are designed by low-power low leakage TSMC SRAM cells. A total of 8KB of memory is required for the full system as data memory. 2.1875KB distributed memory is required for the multiplication operation. All the blocks have clock gating circuit to gate themselves when they are not individually active. For example, when the multiplication block is not active, it is clock-gated. This technique helps us to reduce leakage power by a significant amount and helps us to improve energy efficiency.    

\subsection{Other Components}
Throughout this section special attention was given to the design of the multiplier, sampler and system memory. The implementation of SHA3/SHAKE and the micro-instructions used to control the processor were also discussed. In this subsection the remaining necessary blocks to perform all Saber operations shown in Figure~\ref{full_architecture_saber}(a) are briefly described.

The modules AddPack, Unpack and AddRound perform very similar coefficient-wise operations on the polynomials. Particularly, AddRound performs the addition of a constant followed by the rounding operation which is used during key generation and encryption (see line~5 in Algorithms~\ref{alg:KeyGenPKE}~and~\ref{alg:Enc}). {\rd AddPack} performs a very similar operation during encryption except that it also adds the message encoded (see line~7 of Algorithm~\ref{alg:Enc}). Unpack performs a subtraction also followed by the addition of a constant followed by rounding but the constants used are different (see line~2 of Algorithm~\ref{alg:Dec}). These three operations are straightforward to implement with a single adder and a buffer for the rounding operation. Since these operations are performed coefficient-wise, we parallelize them for four coefficients at a time according to the data width used in the rest of the processor.

The block Verify compares two arrays, i.e., two regions of memory. It is implemented to run in constant time for a given data length. The data is compared using a xor operation and the result of the comparison is accumulated.
The block CMOV implements a secure conditional move which is used during the decapsulation (see lines~4-7 of Algorithm~\ref{alg:Decaps}). The data move takes place according to a given flag, but the block runs in any case to avoid leakage on decryption failures.

The multiplier accepts polynomials that are packed and stored in memory as arrays of $4$ bits if it is a secret polynomial, as arrays of $13$ bits if it is a polynomial in Saber ring, or as arrays of $16$ bits for the rest of polynomials. Internally in the multiplier, all coefficients are fetched as $16$ bit coefficients using the decoders shown in Figure~\ref{full_architecture_saber}(b) and explained in Section~\ref{hw:mult}. However, Saber also contemplates polynomials modulo $p$. The block BS2PolVecP is necessary to transform the packing of polynomials. This operation is not complicated and is implemented using a buffer. Note that all the individual blocks are clock gated when idle to reduce leakage power. 

\subsection{Scalability of Saber ASIC components}
{\rd Saber defines three sets of parameters called LightSaber, Saber, and FireSaber, which match NIST security levels 1, 3, and 5 respectively. All three levels use polynomial degree N= 256, and moduli q= $2^{13}$ and p= $2^{10}$.  The three variants mostly differ in the module dimension, the binomial distribution parameter, and the message space. It should be noted that our Toom-Cook 4-way multiplier supports the multiplication with maximum width, this polynomial multiplication architecture can be reused in any of the Saber variants. However, the implemented ASIC is dedicated to Saber. Hence, the binomial sampler is 4 bits. Binomial sampler architecture slightly changes based on the Saber variant. Due to the lack of reconfigurability of the binomial sampler, this ASIC is dedicated to Saber. However, except this, architecture is fully scalable and support for LightSaber and FireSaber can be added in future version of the IC with minimal change.}

\section{Silicon Results}\label{sec:results}

The integrated circuit is fabricated with 65nm TSMC LP process. 
The wirebond type is chip-on-board and a glob-top encapsulation layer is put on top of the die. However, this area includes test circuits, free space, and memory. It should be noted that we should only care about the accelerator area as data memory can be used as general system memory when the secured connection is already established. Saber core is varied with $V_{DD}$ from 0.7-1.1V and maximum frequency and energy is noted. 

\subsection{Area efficiency and memory footprint comparison}

The total accelerator area is 0.158$mm^2$. It should be noted that the accelerator area does not include area for Keccak as there is no innovation in that and standard Keccak module provided by Keccak team~\cite{keccak_hw} has been used. Keccak is used as Pseudo Random Number Generator (PRNG) which can be generated using other crypto-engine as well~\cite{DBLP:journals/tches/BanerjeeUC19}. {\rd It should be noted that Keccak core takes $0.09mm^2$ active area. However, optimized Keccak or other PRNG might take less area to provide a compact and complete solution. Optimizing Keccak module is beyond the scope of this work. It should be noted that latency and power number includes Keccak module.}

Area requirements for the different blocks have been shown in Figure~\ref{silicon_area}(a). The multiplier is still consuming around ~83.79\% of the area even after multiplier optimization. Other blocks namely binomial sampler, unpack, etc. take around ~2\%. The next biggest block is addround block which consumes 4.17\% of area. The total accelerator area is compared with respect to state-of-the-art PQC cores. A detailed percentage of area is mentioned in figure~\ref{silicon_area}~(a). Comparison with different Saber cores(not silicon-verified) and silicon-verified PQC cores is plotted in Figure~\ref{silicon_area}(b). This work consumes the lowest area to date with respect to PQC cores and accelerators. It should be noted that the lowest area PQC core has been demonstrated by Banerjee et al.~\cite{banerjee20192} though that is done in 40nm hence our accelerator has process disadvantages. The lowest area for Saber accelerator is reported by Zhu et al.~\cite{DBLP:journals/tcasI/ZhuZYZDCWL21} as $0.38mm^2$ which is much higher than we have implemented in our IC as shown in Figure~\ref{silicon_area}(b). This work uses the lowest memory footprint till date. The accelerator needs 10.1875KB to operate which is 4$\times$ lower than existing state-of-the-art as shown in Figure~\ref{silicon_area}(c). State-of-the-art implementation needs 40KB of memory.

\begin{figure}[!ht]
  \centering
  \includegraphics[width=0.45\textwidth]{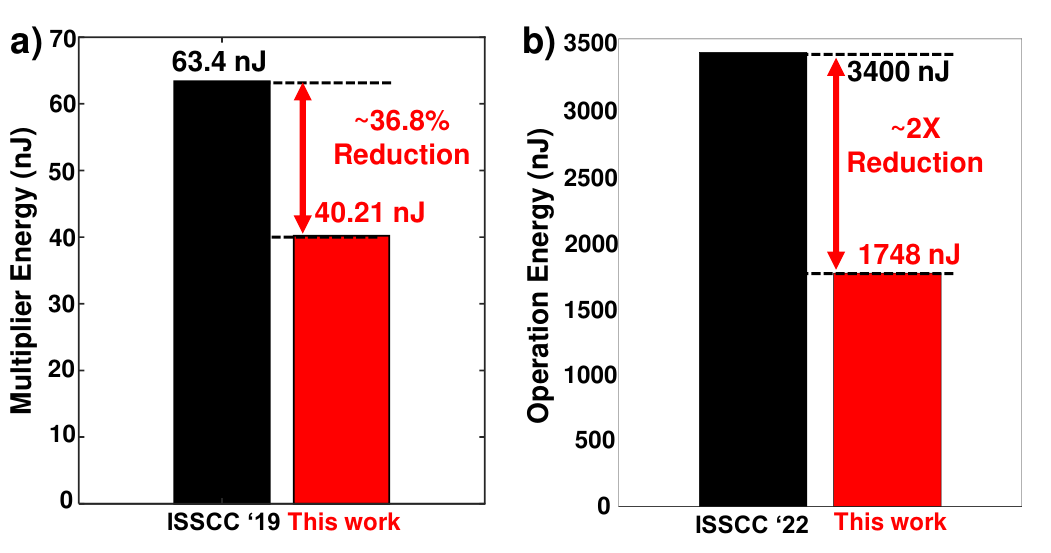}
  \caption{Energy comparison with respect to the state-of-the-art for a) multiplier energy and b) energy per operation.}
  \label{energy_comparison}
\end{figure}

\begin{figure*}[!ht]
  \centering
  \includegraphics[scale=.9]{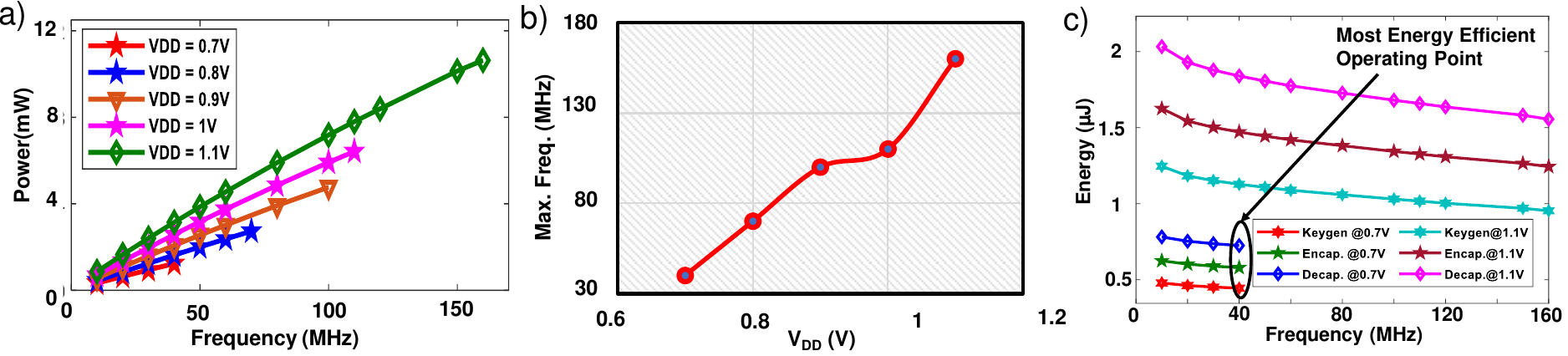}
  \caption{(a) Load characteristics of Saber core (b) Maximum frequency at different $V_{DD}$. (c) The energy at the different operating frequencies. The maximum energy efficiency point is 0.7V, 40MHz.}
  \label{load_characteristics}
\end{figure*}

\subsection{Power \& Energy efficiency}
Figure~\ref{power_comparison} shows the power difference between this work and state-of-the-art implementations. The total power consumed by the design is 333.9uW which is 38\% less that the state-of-the-art PQC core (Figure~\ref{power_comparison}(a)). Moreover, this accelerator is compared with respect to LWRPro \cite{DBLP:journals/tcasI/ZhuZYZDCWL21} which is not silicon verified (Figure~\ref{power_comparison}(b)). LWRPro consumes 39mW which is 118$\times$ higher than this accelerator. {\rd Average power in~\cite{DBLP:journals/iacr/ImranABRP22} varies from 0.855mW to 153.6mW which is higher than our implementation. Also, we refrain from mentioning it as it has incorrect functionality. Another recent solution~\cite{isscc22_agile_processor} provides a reconfigurable architecture for PQC cores. The solution is mostly focused on Kyber, however, can be reconfigured to operate for Saber. This solution mostly focuses on optimizing the latency and consumes 39-368mW (at 0.9V VDD) which is $118\times - 1000\times$ higher than our solution.  } 
Maximum frequency is observed in different $V_{DD}$ in Figure~\ref{load_characteristics}(a). The IC is operational at 40MHz frequency at 0.7V $V_{DD}$. {\rd The accelerator works at the maximum frequency of 160MHz at 1.1V $V_{DD}$. Figure~\ref{load_characteristics}(c) shows the energy of Key generation, encapsulation, and decapsulation by the co-processor. Key generation, encapsulation, and decapsulation take $\le$ 2uJ energy at 1.1V at all frequencies. Energy is reduced as frequency is increased. The increasing frequency will reduce latency, hence the effect of leakage power will be reduced in the final energy calculation. This concept leads to the lowest energy at 0.7V $V_{DD}$ and 40MHz frequency. Key generation, encapsulation, and decapsulation consume 444.1,
579.4 \& 724.5 nJ energy for all the operations combined in the above-mentioned
operating point.} \\
To have an apple-to-apple comparison we compare the energy of point multiplication with Banerjee et al~\cite{banerjee20192}. 
It should be noted that the state-of-the-art PQC core uses NTT structure for point multiplication however we use optimized striding Toom-Cook-based  polynomial multiplication with lazy interpolation which is performance-wise very similar to NTT structure. 
{\rd As discussed in section IIIA earlier, the total latency of the point multiplication is $\frac{64}{n}\times (\frac{n}{2}+n+64+(n-1))$. As $n$ = 4, 1168 clock cycles will be required which is comparable with respect to PQC core published in~\cite{DBLP:journals/tches/BanerjeeUC19} despite not being an NTT-based architecture. Additional 60 clock cycles are required at the evaluation and 70 clock cycles at interpolation. However, it should be noted that interpolation happens once in 3 multiplication. However, to consider the worst case scenario, we need a total of 1298 clock cycles which is similar to NTT architecture presented in ~\cite{banerjee20192}. As our implementation is a low power implementation (333.9uW at 0.7V VDD and 40MHz), this leads to 40.21nJ energy consumption per multiplication which is 36.8\% less than state-of-the-art~\cite{banerjee20192} as shown in Fig.~\ref{energy_comparison}(a). We define a single operation by a combination of key generation, encapsulation, and decapsulation and compare this with respect to state-of-the-art core \cite{isscc22_agile_processor} in FIg.~\ref{energy_comparison}(b). It is observed that the implemented core consumes 1748nJ energy per operation which is ~2$\times$ lower than \cite{isscc22_agile_processor}.}
\begin{table*}[!ht]
  \centering
  \includegraphics[scale=1.2]{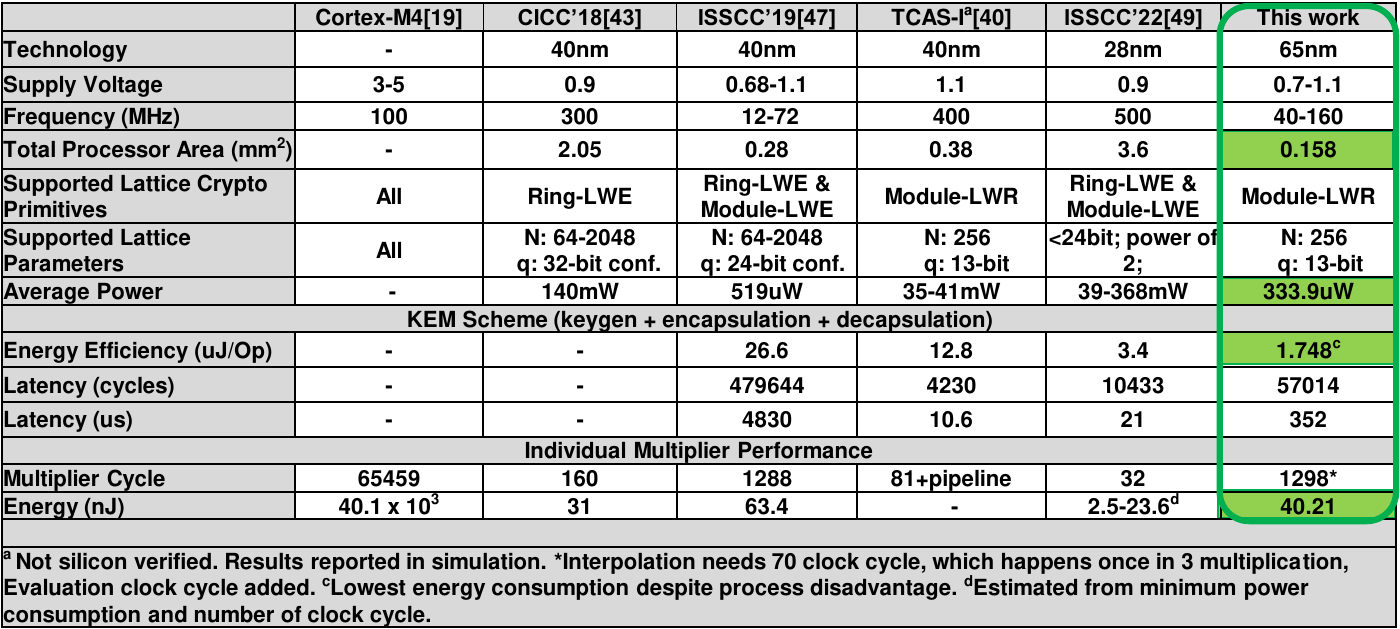}
  \caption{Comparison with state-of-the-art. This accelerator uses the lowest memory footprint and lowest area till date. It also consumes 333.9uW power which is the lowest till now.}
  \label{comparison_table}
\end{table*}

\begin{figure}[!ht]
  \centering
  \includegraphics[scale=.5]{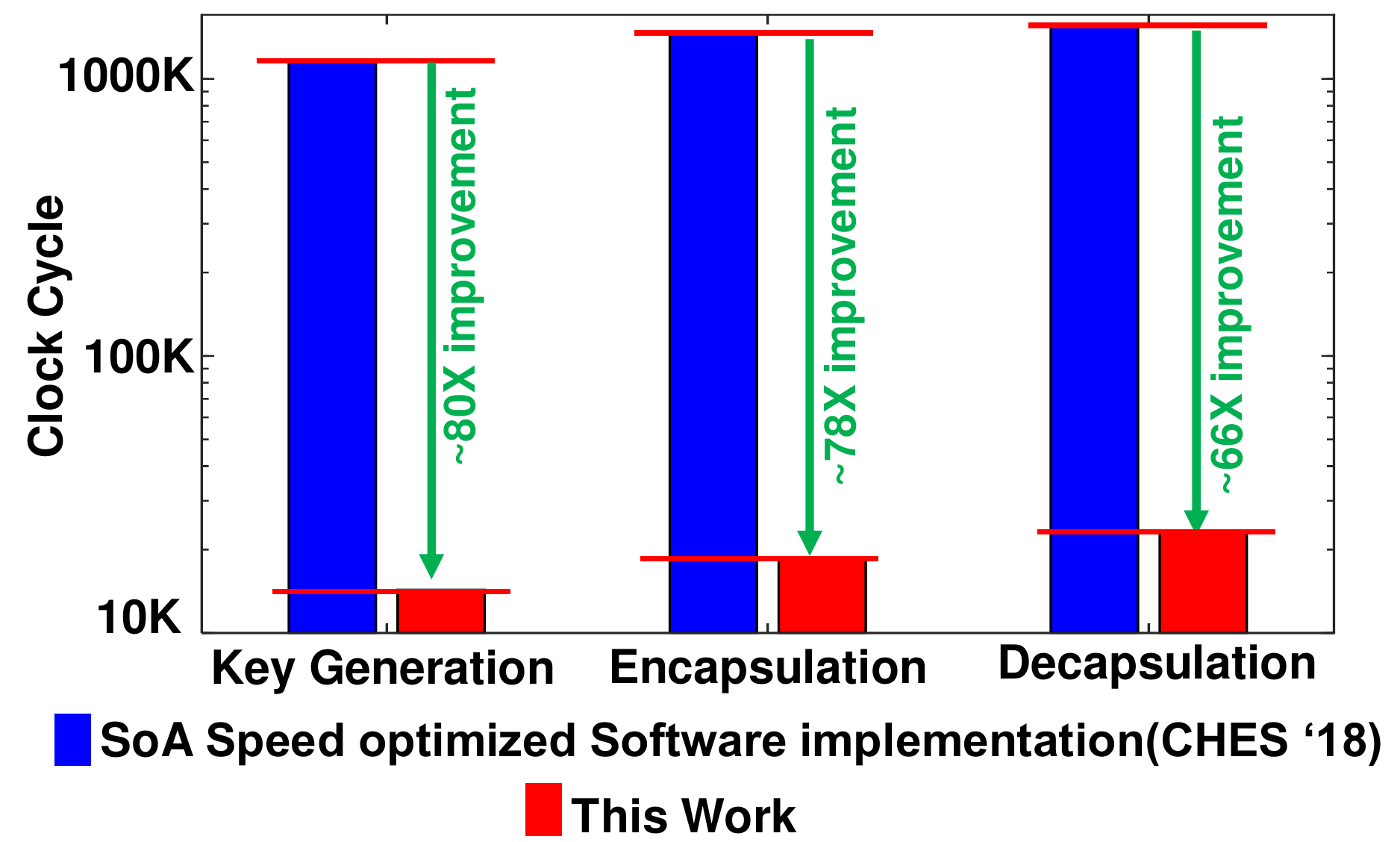}
  \caption{Clock cycle comparison with respect to state-of-the-art speed-optimized software implementation.}
  \label{clock_cycle_compare}
\end{figure}

\subsection{Latency comparison}
Though latency is not our utmost priority as stated earlier, we wanted to match latency with respect to state-of-the post-quantum cores and improve with respect to speed-optimized software implementation. {\rd All KEM operations (keygen, encapsulation, and decapsulation) are observed to be finished within 89, 117 \& 146 $\mu$s respectively at 160MHz frequency (80X improvement over fastest software implementation as shown in Figure~\ref{clock_cycle_compare}). Precisely, the design takes 14642, 18984 \& 23388 clock cycles to finish Saber Keygen, Encapsulation \& Decapsulation respectively.} Note that this design is not optimized based on latency as KEM will be used once in a while to establish secure communication links, unlike symmetric key. However, we still ensure the design is as fast as possible within the scope of a fully compact area and power-optimized design as shown in Figure~\ref{clock_cycle_compare}. 

\section{Conclusion}\label{sec:conclusions}

This work presents the first silicon-verified IC for the Saber accelerator. Moreover, a combined striding Toom-Cook and lazy interpolation Toom-Cook 4-way approach is taken to reduce computation complexity which helps in reducing area and energy overhead which was one of the selection criteria for NIST PQC standardization procedure. Moreover, clock gating in all the blocks including in 3 different memories while other operations are ongoing, multiplication using shift multiplier at evaluation step, less number of loop operations due to algorithmic innovation namely lazy interpolation and optimized memory operation by striding Toom-Cook polynomial multiplication reduces total energy consumption with respect to state-of-the-art PQC core as well requires less memory footprint and area. This IC consumes $0.158mm^2$ area which is the lowest amongst all the PQC cores and accelerators. It also consumes 333.9$\mu$W of average power at the optimum point which is the lowest reported to date amongst the PQC cores. Polynomial multiplication takes 40.21nJ energy with comparable latency with state-of-the-art hardware as shown in Table~\ref{comparison_table}.  


\bibliographystyle{unsrt} {
    \bibliography{main.bib}
}

\end{document}